\documentclass[conference,final,onecolumn,10pt,twoside, narroweqnarray]{IEEEtran}
\usepackage{color, enumitem}
\usepackage{graphicx}
\usepackage{amsmath,amssymb,amsthm}
\usepackage{latexsym}
\usepackage{cite}
\newtheorem{prop}{Proposition}

\newtheorem{lemma}{Lemma}
\newtheorem{theorem}{Theorem}

\usepackage{algorithm}
\usepackage{algorithmic}
\newcommand{\R}{\mathbf{R}}

\newcommand{\bP}{\mathbf{P}}

\newcommand{\bZ}{\mathbf{Z}}
\newcommand{\bR}{\mathbf{R}}
\newcommand{\bV}{\mathbf{V}}
\usepackage{bm}
\usepackage{stfloats}

\ifCLASSINFOpdf
   \graphicspath{{../jpg/}{../pdf/}{../jpeg/}{../eps/}}
   \DeclareGraphicsExtensions{.jpg,.pdf,.jpeg,.png,.eps}
\else
  \DeclareGraphicsExtensions{.eps}
\fi

\ifCLASSOPTIONcompsoc
  \usepackage[caption=false,font=normalsize,labelfont=sf,textfont=sf]{subfig}
\else
  \usepackage[caption=false,font=footnotesize]{subfig}
\fi
\ifCLASSOPTIONcompsoc
\else
\fi

\usepackage{setspace}

\begin{document}
%
\title{Movement-efficient Sensor Deployment\\ in Wireless Sensor Networks\vspace{-10pt}}


\author{Jun Guo and Hamid Jafarkhani\\
Center for Pervasive Communications \& Computing,  University of California, Irvine\\\vspace{-20pt}}

\maketitle

\begin{abstract}
We study a mobile wireless sensor network (MWSN) consisting of multiple mobile sensors or robots.
Two key issues in MWSNs —— energy consumption, which is dominated by sensor movement, and sensing coverage —— have attracted plenty of attention, but the interaction of these issues is not well studied.
To take both sensing coverage and movement energy consumption into consideration, we model the sensor deployment problem as a constrained source coding problem.
Our goal is to find an optimal sensor deployment to maximize the sensing coverage with specific energy constraints.
We derive necessary conditions to the optimal sensor deployment with (i) total energy constraint and (ii) network lifetime constraint. Using these necessary conditions, we design Lloyd-like algorithms to provide a trade-off between sensing coverage and energy consumption.
Simulation results show that our algorithms outperform the existing relocation algorithms.
\end{abstract}


%
\IEEEpeerreviewmaketitle

\section{Introduction}
\label{secIntro}
Deploying multiple nodes, sensors or robots, to monitor the target environment is the primary objective of the mobile wireless sensor networks (MWSNs).
To evaluate the sensing quality, the binary disk coverage model, in which each sensor can only cover a disk with the radius $R_s$, is widely used in MWSNs.
Due to different sensing tasks in the real world, multiple coverage measurement and deployment algorithms are well studied in recent decades, look at \cite{SD} and the references therein.

Four popular coverage categories are (i) area coverage, (ii) target coverage, (iii) barrier coverage, and (iv) evenly deploying the sensors.
A natural sensing task is to maximize the area coverage, which is formulated by the total area covered by sensors.
In another popular coverage task, target coverage, the specific target locations are detected and reported by static sensors.
In this case, sensors or robots are required to collect detailed information from the discrete targets.
A full-target coverage is achieved if and only if every discrete target in the 2-dimensional region is covered by at least one sensor.
In another popular coverage task, barrier coverage, sensors are moving along the boundary to detect intruders as they cross the border of a region or domain.
To obtain full-barrier coverage, one should place sensors to cover the whole barrier or boundary.
Finally, an even deployment of the sensors requires them to form a Centroidal Voronoi Tessellation (CVT).
It is mainly used when there is no specific target.
The widely used CVT model (see more details in Section \ref{sec:model}) is a quantizer where its distortion is the sensing uncertainty \cite{SD,YBZ,GJ,VD}.

Energy efficiency is another key issue in MWSNs as most sensors have limited battery energy, and it is inconvenient or even unfeasible to replenish the batteries of numerous densely deployed sensors.
In general, the energy consumption of a device includes communication energy, data processing energy, sensing energy, and movement energy.
In fact, sensor movement has a much higher energy consumption compared to other types of energy \cite{KD,JW}, and then dominates the energy consumption.
Guiling et al. \cite{GW} study the optimal angular velocity and the optimal acceleration to minimize the energy consumption for motion.
Simulation results in \cite{GW} show that , the energy consumption for motion with the optimal angular velocity setting is approximately linear to the movement distance.
In fact, the linear movement energy consumption is a popular assumption and widely adopted in the literature \cite{WL,KS,SP,AJ,ZL,ZL2,AT,ZFG,AWAE,DYJH,SH,ZB,ABJ,XBCHW,DD,HLC,YM,YK,ZW,SMJ}.
Particularly, the movement energy consumption in some specific sensors is 5.976J/m \cite{YM}.

A huge body of literature exists on reducing movement energy consumption with coverage guarantee.
First, the minimization of energy consumption or movement distance with the full-area coverage guarantee is well studied by \cite{WL,KS,SP,AJ,YK,ZW,SMJ}.
In \cite{WL}, the author applies Hungarian Algorithm to minimize the total energy consumption after the full-area coverage is achieved by Genetic Algorithm.
Similarly, the grid-based algorithms are proposed in \cite{SP} to reduce the total moving distance while keeping the full-area coverage and full-connectivity.
Kuei-Ping et al. \cite{KS} propose a distributed partition avoidance lazy movement (PALM) protocol, which avoids unnecessary movement, to ensure both area coverage and connectivity.
Shuhui et al. \cite{SMJ} provide a scan-based relocation algorithm, SAMRT, which is supposed to be energy-effective with densely deployed sensors.
Three virtual force based algorithms, VFA \cite{YK}, DSSA \cite{ZW}, and HEAL \cite{AJ} are proposed to maximize the area coverage while saving energy.
In \cite{YK}, the authors attempt to prolong the network lifetime by (a) disabling any virtual forces on a sensor whenever the current distance reaches the distance limit and (b) keeping track of the maximum coverage.
In \cite{ZW}, the authors put the local sensor density into the virtual force calculation, and thus avoid unnecessary movements in the region with densely deployed sensors.
In \cite{AJ}, HEAL is designed to mend area coverage holes while minimizing the moving distance. However, its prerequisite, that there are enough sensors to achieve full-coverage, limits HEAL's usage.

Furthermore, the energy-efficient mobile sensor relocation with CVT guarantee has been studied in recent years.
A natural approach is to add a penalty term, which is related to the moving distance, into the objective function.
In \cite{YBZ}, the authors propose two algorithms, Lloyd-$\alpha$ and DEED, to implement CVT with a movement related penalty function.
For Lloyd-$\alpha$, the movement in each iteration is scaled by a parameter $\alpha\in[0,1]$.
In DEED, the penalty function is properly selected with a positive definite matrix, and then the movement is optimized with the help of gradient and Hessian matrix of the distortion.

Note that the existing energy minimization and network lifetime maximization methods are based on the premise of full-coverage or CVT. However, full-coverage or CVT is unnecessary and even infeasible in some applications, especially when sensors are not abundant enough to cover the target regions.
For example, to estimate the total amount of a rainfall with energy limited sensors, full-area coverage and CVT are unnecessary and one should pay more attention to the energy consumption to prolong the network lifetime.
Furthermore, the existing solutions cannot maximize the coverage or CVT with the specific total energy or network lifetime constraints.
While there has been extensive work on total energy minimization or network lifetime maximization with full-coverage, to the best of our knowledge, the real trade-off between (generalized) coverage and movement energy consumption in MWSNs has not been considered in the literature.

In this paper, we study the sensor deployment problem in MWSNs and make the following contributions:
(1) Taking the total energy consumption and network lifetime as two separate constraints, we consider two constrained optimization problems for optimal sensor deployment.
(2) We provide the necessary conditions for the above constrained optimization problems.
(3) We also design centralized Lloyd-like algorithms to optimize the sensor deployment with the total energy and network lifetime constraints.

The rest of this paper is organized as follows: In Section \ref{sec:model}, we introduce the sensing performance model, and show that coverage problems can be converted to a CVT problem with the properly selected density function.
In Section \ref{sec:ProblemA}, we study sensor deployments for the MWSNs in which the total energy consumption is constrained.
In Section \ref{sec:ProblemB}, we discuss sensor deployments for the MWSNs in which the individual energy consumptions are constrained (or a network lifetime is constrained).
In Section \ref{sec:simulation}, we present numerical simulations.
In Section \ref{sec:conclusion}, we draw our main conclusions, and discuss the extensions of our approachs to the target coverage and barrier coverage tasks.
\section{Sensing Performance without Energy Constraint}\label{sec:model}
Let $\Omega$ be a simple convex polygon in $\Re^2$ including its interior.
Given $N$ sensors in the target area $\Omega$, sensor deployment before and after the relocation
are, respectively, defined by $\tilde{\bP} = (\tilde{p}_1, \dots, \tilde{p}_N)\subset\Omega^{N}$ and $\bP = (p_1, \dots, p_N)\subset\Omega^{N}$, where $\tilde{p}_n$ is Sensor $n$'s initial location and $p_n$ is Sensor $n$'s final location.
Let $\mathcal{I}_{\Omega}=\{1,\dots,N\}$ be the whole set of sensors in the WSN.
For any point $w\in\Omega$, the density function $f(w)$ reflects the importance of an event at point $w$.
A cell partition $\bR(\bP)$ of $\Omega$ is a collection of disjoint subsets of $\{R_n(\bP)\}_{n\in\mathcal{I}_{\Omega}}$ whose union is $\Omega$.
We assume that Sensor $n$ only monitors the events that occurred in its cell partition $R_n(\bP)$, $\forall n\in\mathcal{I}_{\Omega}$.
Let $\|\cdot\|$ denote the Euclidean distance, and $B(c,r)=\{w|\ \|w-c\|\le r\}$ be a disk centered at $c$ with radius $r$.

To evaluate the sensing uncertainty in heterogeneous WSNs, we consider the Centroidal Vonoroi Tessellation function \cite{SD,GJ,YBZ,VD} defined as
\vspace{-5pt}
\begin{equation}
\small
D(\bP)=\sum_{n=1}^{N}\int_{R_n(\bP)}\eta_n\|p_n-w\|^2f(w)dw,
\label{distortion}
\vspace{-3pt}
\end{equation}
where the sensing cost parameters ${\eta_n\in(0,1]}$ is a constant
that depends on Sensor $n$'s characteristics and $f(w)$ is a density function that reflects the target importance.
In homogeneous MWSNs, sensors have identical parameters, i.e., $\!\eta_{\!n}\!\!=\!\!1,\forall\!n\!\in\!I_{\!\Omega}$.

The optimal partition for the performance function (\ref{distortion}) is Multiplicatively Weighted Voronoi Diagram (MWVD) \cite{GJ}, which can be applied to both homogeneous and heterogeneous WSNs.
The MWVD of $\Omega$ generated by $\bP$ is the collection of sets $\{V_n(\bP)\}_{n\in I_{\Omega}}$ defined by
\begin{equation}
V_n\!(\bP)\!=\!\{w\!\in\!\Omega|\eta_n\|w\!-\!p_n\|^2\!\le\!\eta_m\|w\!-\!p_m\|^2,\forall m\!\in\!I_{\Omega}\}.
\label{MWVD}
\end{equation}
In particular, the MWVD for homogeneous WSNs degenerates to the Voronoi Diagram \cite{VD}.
From now on, we use
$\bV(\bP)=\{V_n(\bP)\}_{n\in\mathcal{I}_{\Omega}}$ to replace partition $\bR(\bP)=\{R_n(\bP)\}_{n\in\mathcal{I}_{\Omega}}$.

\section{The Sensor Deployment with a Total Energy Constraint}\label{sec:ProblemA}
\subsection{Problem formulation}
In this section, we review a classic energy consumption model for the mobile sensor networks.
Since the sensor movement dominates the power consumption, we only consider the power consumption for sensor movement.
As we mentioned in Section \ref{secIntro}, the energy consumption for movement is linear to the moving distance.
Therefore, the energy consumption for Sensor $n$ can be defined as
\vspace{-2pt}
\begin{equation}
E_n(\bP) = \xi_n\|p_n-\tilde{p}_n\|,
\label{individualPower}
\vspace{-2pt}
\end{equation}
where the moving cost parameter $\xi_n$ is a predetermined constant that depends on Sensor $n$'s energy efficiency and $p_n$ is Sensor $n$'s destination.
Accordingly, the total power consumption is
\vspace{-2pt}
\begin{equation}
E(\bP) = \sum_{n=1}^{N}E_n(\bP) = \sum_{n=1}^{N}\xi_n\|p_n-\tilde{p}_n\|.
\label{totalPower}
\vspace{-2pt}
\end{equation}
Now, we are ready to propose the main goal: minimizing the sensing uncertainty defined by (\ref{distortion}) given a constraint on the total energy consumption defined by (\ref{totalPower}).
The constrained optimization problem $\mathcal{A}$ is defined as
\begin{align}
& \underset{\bP}{\text{minimize}} \;\;\;\; D(\bP) \\
& \text{~~~~s.t.  } \;\;\;\;\;\;\; E(\bP)\leq\gamma
\label{totalConstraint}
\end{align}
where $\gamma$ is the maximum energy consumption.

\subsection{The Optimal Sensor Deployment}\label{sec:opt1}
Before going through the details about the optimal sensor deployment with a total energy constraint, we discuss the optimal partition.
Sensing uncertainty is determined by both sensor movement and cell partition, but energy consumption only depends on the sensor movement.
In other words, MWVDs are also the optimal partitions of the MWSNs with a constrained total energy consumption.
Now, we will discuss the optimal sensor deployment that minimizes (\ref{distortion}) with energy constraints.
\begin{lemma}
Let $\bP^*=(p^*_1,\dots,p^*_N)$ be the optimal deployment in MWSNs with a total energy consumption.
Sensor $n$'s optimal location $p^*_n$ is placed between its initial location and the geometric centroid of its MWVD, i.e.,
$p^*_n=\delta \tilde{p}_n + (1-\delta)c_n(P^*)$,
where $\delta\in[0,1]$ and $c_n(P^*)\!\!=\!\!\frac{\int_{V\!_n(\bP^*)}wf(w)dw}{\int_{V\!_n(\bP^*)}f(w)dw}$ is the geometric centroid of $V_n(\bP^*)$.
\label{oninterval}
\end{lemma}
\vspace{-5pt}
The proof is provided in Appendix \ref{appendixL1}.

Lemma \ref{oninterval} shows that the geometric centroid still plays an important role in the optimal deployment with a total energy consumption.
The main change is that instead of moving the sensor to the geometric centroid, we should move it towards the geometric centroid but stop before reaching the centroid.
Next, we introduce several important concepts to calculate the optimal $\delta$ values.
First, sensors in MWSNs can be classified according to the moving distance into (a) dynamic sensors who have positive moving distance and (b) static sensors who stand still.
Let $\mathcal{I}_{d}(\bP)=\{n|\|p_n-\tilde{p}_n\|>0,n\in\mathcal{I}_{\Omega}\}$ and $\mathcal{I}_{s}(\bP)=\{n|p_n=\tilde{p}_n,n\in\mathcal{I}_{\Omega}\}$ be, respectively, the dynamic sensor set and static sensor set.
Similar to the definition in \cite{GJ}, let $c_n(\bP)=\frac{\int_{V_n(\bP)}p_{\!n}f(w)dw}{\int_{V_n(\bP)}f(w)dw}$ and $v_n(\bP)=\int_{V_n(\bP)}f(w)dw$, be respectively, the geometric centroid and the volume of $V_n(\bP)$.
Another concept, moving efficiency, is defined as $\rho_n(\bP)=\frac{\varrho_n(\bP)}{\varsigma_n(\bP)}$ to reflect Sensor $n$'s ability to decrease distortion by movement, where $\varsigma_n(\bP)\!=\!\frac{\xi_n^2}{\eta_nv_n(\bP)}$ and $\varrho_n(\bP)\!=\!\xi_n\|p_n\!-\!c_n(\bP)\|$.
In addition, let $\bar{\rho}(\bP)\!=\!\frac{\left(\sum_{i\in\mathcal{I}_{d}(\bP)}\xi_i\|\Gamma_i(\bP)\|\right)\!-\!\gamma}{\sum_{i\in\mathcal{I}_{d}(\bP)}\varsigma_i(\bP)}$ be a moving efficiency threshold, where $\Gamma_n(\bP)=c_n(\bP)-\tilde{p}_n$ is the vector from Sensor $n$'s initial location to the geometric centroid of $V_n(\bP)$.
\vspace{-5pt}
\begin{prop}
Let $\bP^*\!\!=\!\!(p^*_1,\!\dots\!,p^*_N)$ be the optimal sensor deployment in MWSNs with a total energy consumption $\gamma\ge0$.
When $\sum_{n=1}^N\xi_n\|\Gamma_n(\bP^*)\|\le\gamma$, the necessary condition for the optimal deployment is $p^*_n=c_n(\bP^*)$, $\forall n\in\mathcal{I}_{\Omega}$.
Otherwise, the necessary conditions for the optimal deployment are:\vspace{5pt}\\
(i)
$\rho_i(\bP^*)\!=\!\bar{\rho}(\bP^*)\!\ge\rho_j(\bP^*)$, $\forall i\!\in\!\mathcal{I}\!_{d}(\bP^*), j\!\in\!\mathcal{I}\!_{s}(\bP^*)$;\vspace{5pt}\\
(ii) $p^*_n\!=\!c_n(\bP^*\!)\!-\!\frac{\varsigma_n(\bP^*)\bar{\rho}(\bP^*)\Gamma_n\!(\!\bP^*\!)\!}{\xi_n\|\Gamma_n\!(\bP^*)\|}$, $\forall n\!\in\!\mathcal{I}\!_{d}(\bP^*)$;
\label{PropA}
\end{prop}
The proof is provided in Appendix \ref{appendixP1}.

Note that $p^*_n=c_n(\bP^*)$ is necessary for the optimal sensor deployment without constraints.
When $\gamma$ is large enough, indicating a loose constraint, Lloyd Algorithm, without
considering the energy constraint, can be used to find the optimal sensor deployment without considering the energy constraint.
On the other hand, when $\gamma$ is small, the energy constraint plays an important role in the sensing task.
Condition (i) reveals the basic principle of sensor division: sensors who have large moving efficiencies are more likely to be selected as dynamic sensors, and Condition (ii) indicates the optimal moving directions and moving distances for the dynamic sensors.
With the help of the necessary conditions in Proposition 1, a Lloyd-based algorithm, Efficient Movement Lloyd (EML) Algorithm, is proposed to find the optimal sensor deployment in the next subsection.
\subsection{Efficient Movement Lloyd Algorithm}\label{sec:EML}
Before we discuss the details of EML Algorithm, we introduce the important concept of local distortion.
The global distortion is a summation of $N$ local distortions defined by
\begin{equation}
D_n(\bP) = \int_{R_n(\bP)}\eta_n\|p_n-w\|^2f(w)dw, n\in\mathcal{I}_{\Omega}.
\label{localdistortion}
\end{equation}
For simplicity, let $\Gamma_n(\bP)=c_n(\bP)-\tilde{p}_n$ be the vector from the initial location to the geometric centroid.
Note that the auxiliary variables $\Gamma_n(\bP)$, $\varrho(\bP)$, and $\varsigma(\bP)$ depend on MWVDs $\{V_n(\bP)\}_{n\in\mathcal{I}_{\Omega}}$, which are also functions of $\bP$.
Therefore, multiplicatively weighted Voronoi partition \cite{VD} is a necessary step before calculating the above auxiliary variables.
Now we introduce the EML Algorithm.
EML Algorithm iterates between two steps: (1) Fixing the sensor deployment and optimizing the partition: Partitioning is done by assigning MWVDs to each sensor node; (2) Fixing the partition and optimizing the sensor deployment: If $\gamma\ge\sum_{n=1}^{N}\xi_n\Gamma_n(\bP)$, Sensors move to their geometric centroid.
Otherwise, Sensors move to the locations $p_n=\tilde{p}_n+\left(1-\frac{\varsigma(\bP)\bar{\rho}(\bP)}{\xi_n\|\Gamma_{\!\!n}(\bP)\|}\right)\Gamma_n(\bP)$.
Sensor $n$'s moving distance $\mathcal{M}_n$ is determined by an iterative algorithm.
More details about EML Algorithm are shown in Algorithm \ref{EMLA}. (Note that $\mathcal{I}_{d}$ and $\mathcal{I}_{s}$ are  pre-determined dynamic and static sensor sets without information about sensor deployment $\bP$.)

\begin{algorithm}[!htb]
\caption{Efficient Movement Lloyd Algorithm in heterogeneous WSNs}
\label{EMLA}
\begin{algorithmic}[1]
\REQUIRE~~\\
Target area $\Omega$\\
Probability density function $f(\cdot)$\\
The number of sensor nodes: $N$\\
The number of total iterations $Iter_{max}$\\
The initial sensor deployment $\tilde{\bP}$\\
The energy constraint $\gamma$\\
\ENSURE~~\\
Final sensors deployment $\bP$\\
Distortion $D(\bP)$\\
\STATE Generate initial locations for sensor nodes
\FOR{$iter$ = 1 to $Iter_{max}$}
\STATE Do multiplicatively weighted Voronoi partition
\STATE Calculate $\{\Gamma_{\!\!n}\!(\bP\!)\}_{n\in\mathcal{I}_{\Omega}}$ and $\{\varsigma_n\!(\bP\!)\}_{n\in\mathcal{I}_{\Omega}}$
\IF{$\gamma\ge\sum_{n=1}^{N}\xi_n\Gamma_n(\bP)$}
\STATE Update movement $\mathcal{M}_n=\Gamma_n(\bP)$\\
\ELSE
\STATE Set $\mathcal{I}_{d}=\mathcal{I}_{\Omega}$ and $\mathcal{M}_n=0, \forall n\in\mathcal{I}_{\Omega}$
\STATE Calculate $z_n\!=\!\xi_n\|\Gamma_n(\bP)\|\!-\!\varsigma_n\!(\bP)\frac{\left[\sum_{i\in\mathcal{I}_{d}}\xi_n\|\Gamma_i(\bP)\|\right]-\gamma}{\sum_{i\in\mathcal{I}_{d}}\varsigma_i(\bP)} , n\in\mathcal{I}_{d}$
\WHILE{$\exists n\in\mathcal{I}_{d}$ such that $z_n\le0$}
\STATE Update $\mathcal{I}_{d}=\mathcal{I}_{d}-\bigcup_{z_n\le0}n$
\STATE Update $\{z_n\}_{n\in\mathcal{I}_{d}}$
\ENDWHILE
\STATE Update moving efficiency threshold $\bar{\rho}(\bP)\!=\!\frac{\left(\sum_{i\in\mathcal{I}_{d}(\bP)}\xi_i\|\Gamma_i(\bP)\|\right)\!-\!\gamma}{\sum_{i\in\mathcal{I}_{d}(\bP)}\varsigma_i(\bP)}$
\STATE Update movement $\mathcal{M}_n=\left(1-\frac{\varsigma_n(\bP)\bar{\rho}(\bP)}{\xi_n\|\Gamma_{\!\!n}(\bP)\|}\right)\Gamma_n(\bP), \forall n\in\mathcal{I}_d$
\ENDIF
\STATE Update sensor deployment $p_n=\tilde{p}_n+\mathcal{M}_n$
\ENDFOR
\end{algorithmic}
\end{algorithm}
Next, we show that EML Algorithm is an iterative improvement algorithm and the distortion converges.
\begin{theorem}
EML Algorithm is an iterative improvement algorithm, i.e., the distortion decreases in each iteration, and its distortion converges.
\label{T1}
\end{theorem}
The proof is provided in Appendix \ref{appendixT1}.
\section{The sensor deployment with a network lifetime constraint}\label{sec:ProblemB}
\subsection{Problem formulation}
Another important objective could be minimizing the sensing uncertainty defined by (\ref{distortion}) given a constraint on the network lifetime.
Let $e_n$ be the residual energy on Sensor $n$.
To ensure the network lifetime, $T$, we should have
\begin{equation}
\max_n\left(e_n-E_n(\bP)\right)\ge \alpha T,
\end{equation}
where $\alpha$ is the power consumption for Sensor $n$ after the sensor relocation and $E_n(\bP)$ is the individual energy consumption defined by (\ref{individualPower}).
Therefore, one can achieve the network lifetime, $T$, by properly setting the maximum individual energy consumption as $\gamma_n=e-\alpha T, n\in I_{\Omega}$.
Now, we are ready to define the main goal: minimizing the sensing uncertainty defined by (\ref{distortion}) given constraints on the individual energy consumption defined by (\ref{individualPower}).
The constrained optimization problem $\mathcal{B}$ is thus
\begin{align}
& \underset{\bP}{\text{minimize}} \;\;\;\;\;\;\;\;\;\;\;\; D(\bP) \\
& \text{~~~~s.t.} \;\;\;\;\; E_n(\bP)\leq\gamma_n, n\in I_{\Omega},
\label{individualConstraint}
\end{align}
where $\gamma_n$ is the maximum individual energy consumption on Sensor $n$.
\subsection{The Optimal Sensor Deployment}\label{sec:opt2}
\begin{prop}
Let $\tilde{\bP}\!\!=\!\!(\tilde{p}_1,\!\dots\!,\tilde{p}_N)$ be the initial sensor deployment.
The necessary conditions for the optimal deployments $\bP^*\!\!=\!\!(p^*_1,\!\dots\!,p^*_N)$ in a MWSN with performance function (\ref{distortion}) and constraint (\ref{individualConstraint}) is
\begin{align}
p^*_n\!=\!\tilde{p}_n\!+\!\min\!\left(1,\frac{\gamma_n}{\xi_n\|\Gamma_{\!n}\!(\bP^*)\|}\right)\!\Gamma_{\!n}\!(\bP^*), n\in I_{\Omega},
\label{optP}
\end{align}
where $\Gamma_n(\bP^*)=c_n(\bP^*)-\tilde{p}_n$ and $c_n(\bP^*)=\frac{\int_{V_n(\bP^*\!)}wf(w)dw}{\int_{V_n(\bP^*\!)}f(w)dw}$ is the geometric centroid of the MWVD $V_n(\bP^*)$.
\label{PropB}
\end{prop}
The proof is provided in Appendix \ref{appendixP2}.

Unlike case of the total energy constraint in Section \ref{sec:ProblemA}, every sensor with limited individual energy should move towards the geometric centroid of its cell partition.
With the help of the necessary conditions in Proposition \ref{PropB}, we design  Constrained Movement Lloyd (CML) Algorithm to find the optimal sensor deployment with individual energy constraints in the next subsection.
\subsection{Constrained Movement Lloyd Algorithm}\label{sec:CML}
The feasible region $F_n(\tilde{\bP})$ for Sensor $n$ is defined as a disk centered at $\tilde{p}_n$ with the radius of $r_n=\frac{\gamma_n}{\xi_n}$.
To avoid exceeding the maximum individual energy, sensors can only move to the destinations within the feasible regions $F_n(\tilde{\bP})$.
Like Lloyd Algorithm, Constrained Movement Lloyd (CML) Algorithm consists of two steps: (1) The optimal partition: Partitioning is done by assigning the MWVDs to each sensor node; (2) Local optimization: sensors move to their new locations $p_n=\tilde{p}_n+\min\!\left(1,\frac{\gamma_n}{\xi_n\|\Gamma_{\!\!n}\!(\bP)\|}\right)\!\Gamma_{\!n}\!(\bP), \forall n\in\mathcal{I}_{\Omega}$.
More details about CML Algorithm are shown in Algorithm \ref{CMLA}.

\begin{algorithm}[!htb]
\setlength\abovecaptionskip{0pt}
\setlength\belowcaptionskip{-10pt}
\caption{Constrained Movement Lloyd Algorithm in heterogeneous WSNs}
\label{CMLA}
\begin{algorithmic}[1]
\REQUIRE~~\\
Target area $\Omega$\\
Probability density function $f(\cdot)$\\
The number of sensor nodes: $N$\\
The number of total iterations $Iter_{max}$\\
The initial sensor deployment $\tilde{\bP}$\\
The energy constraints $\{\gamma_n\}_{n\in\mathcal{I}_{\Omega}}$\\
\ENSURE~~\\
Sensors deployment $\bP$\\
Distortion $D(\bP)$\\
\STATE Generate initial locations for sensor nodes
\FOR{$iter$ = 1 to $Iter_{max}$}
\STATE Do multiplicatively weighted Voronoi partition
\STATE Calculate $\{\Gamma_n(\bP)\}_{n\in\mathcal{I}_{\Omega}}$
\FOR{$n$ = 1 to $N$}
\STATE Update movement $\mathcal{M}_n\!=\!\min\!\left(1,\frac{\gamma_n}{\xi_n\|\Gamma_{\!\!n}\!(\bP)\|}\right)\Gamma_{\!\!n}\!(\bP)$
\ENDFOR
\STATE Update sensor deployment $p_n=\tilde{p}_n+\mathcal{M}_n$
\ENDFOR
\end{algorithmic}
\end{algorithm}
\vspace{-10pt}
\begin{theorem}
CML Algorithm is an iterative improvement algorithm, i.e., the distortion decreases in each iteration, and its distortion converges.
\label{T2}
\end{theorem}
\vspace{-5pt}
The proof is provided in Appendix \ref{appendixT2}.
\section{Performance Evaluation}\label{sec:simulation}
We provide the simulation results for two different MWSNs: (1) MWSN1: A homogeneous MWSN in which all sensors have the same parameter.
(2) MWSN2: A heterogeneous MWSN including two types of sensors: eight strong sensors and twenty-four weak sensors.
The sensing and moving cost parameters are set in the Table \ref{SPT}.
Note that the sensing radius is $\frac{R_s}{\sqrt{\eta_n}}$ rather than $R_s$.
In addition, we generate initial sensor deployments on $\Omega$ randomly, i.e., every node location is generated with uniform distribution on $\Omega$.
The maximum number of iterations is set to 100.

\vspace{0pt}
\begin{table}[!htb]
\setlength\abovecaptionskip{0pt}
\setlength\belowcaptionskip{-28pt}
\centering
\caption{Simulation Parameters}
\begin{tabular}{|c|c|c|c|c|c|c|}
\hline
Parameters & $\!N\!$ & $\!\!\eta_{1}\!-\!\eta_{8}\!\!$ & $\!\!\eta_9\!-\!\eta_{32}\!\!$ & $\!\!\xi_1\!-\!\xi_8\!\!$ & $\!\!\xi_9\!-\!\xi_{32}\!\!$ & $\!\!R_s\!\!$\\
\hline
MWSN1 & 32 & 1 & 1 & 1 & 1 & 0.2\\
\hline
MSWN2 & 32 & 1 & 2 & 3 & 1 & 0.3\\
\hline
\end{tabular}
\label{SPT}
\end{table}
\vspace{0pt}

To evaluate the performance of EML and CML algorithms, we compare (1) the total moving distance, and (2) the maximum individual moving distance with GH \cite{WL}, VFA \cite{YK}, Lloyd-$\alpha$ \cite{YBZ}, and DEED \cite{YBZ}.
Fig \ref{EMLA} illustrates examples of the final deployments of EML and CML algorithms in both MWSN1 and MWSN2.
The total moving distance constraint in both MWSN1 and MWSN2 is set to $\gamma=8$.
The individual moving distance constraints in MWSN1 and MWSN2 are, respectively, set to $0.4$ and $1$.
After running the EML algorithm, 30 sensors in MWSN1 and 28 sensors in MWSN2 are dynamic.
The coverage area in MWSN1 is increased from 0.53 to 0.77.
The coverage area in MWSN2 is increased from 0.54 to 0.71.
After running the CML algorithm, every sensor in MWSN1 and MWSN2 has a positive moving distance.

We also compare the area coverage of our algorithm with that of the other deployment algorithms.
The area coverage is defined as the proportion of the target that covered by at least one sensor.
Therefore, area coverage $C^{\mathcal A}(\bP)$ is formulated as
\vspace{0pt}
\begin{equation}
C^{\!\mathcal A}\!(\!\bP)\!\!=\!\!\frac{\int_{\bigcup_{n\!=\!1}^N\!B\left(\!p_n,\frac{R_s}{\sqrt{\eta_n}}\!\right)\!}\!dw}{\int_{\Omega}dw}\!\!=\!\!\frac{\sum_{n\!=\!1}^{N}\!\int_{V\!_n\!(\bP)\!\bigcap\! B\!\left(\!p_n, \frac{R_s}{\sqrt{\eta_n}}\!\right)}\!\!dw}{\sum_{n\!=\!1}^{N}\int_{V\!_n(\bP)}\!dw},
\vspace{0pt}
\end{equation}
where $\frac{R_s}{\sqrt{\eta_n}}$ is Sensor $n$'s sensing radius.
Intuitively, to decrease the sensing uncertainty, the sensors will be evenly distributed, and then provide a large area coverage.

\begin{figure}[!t]
\centering
\subfloat[]{\includegraphics[width=3in]{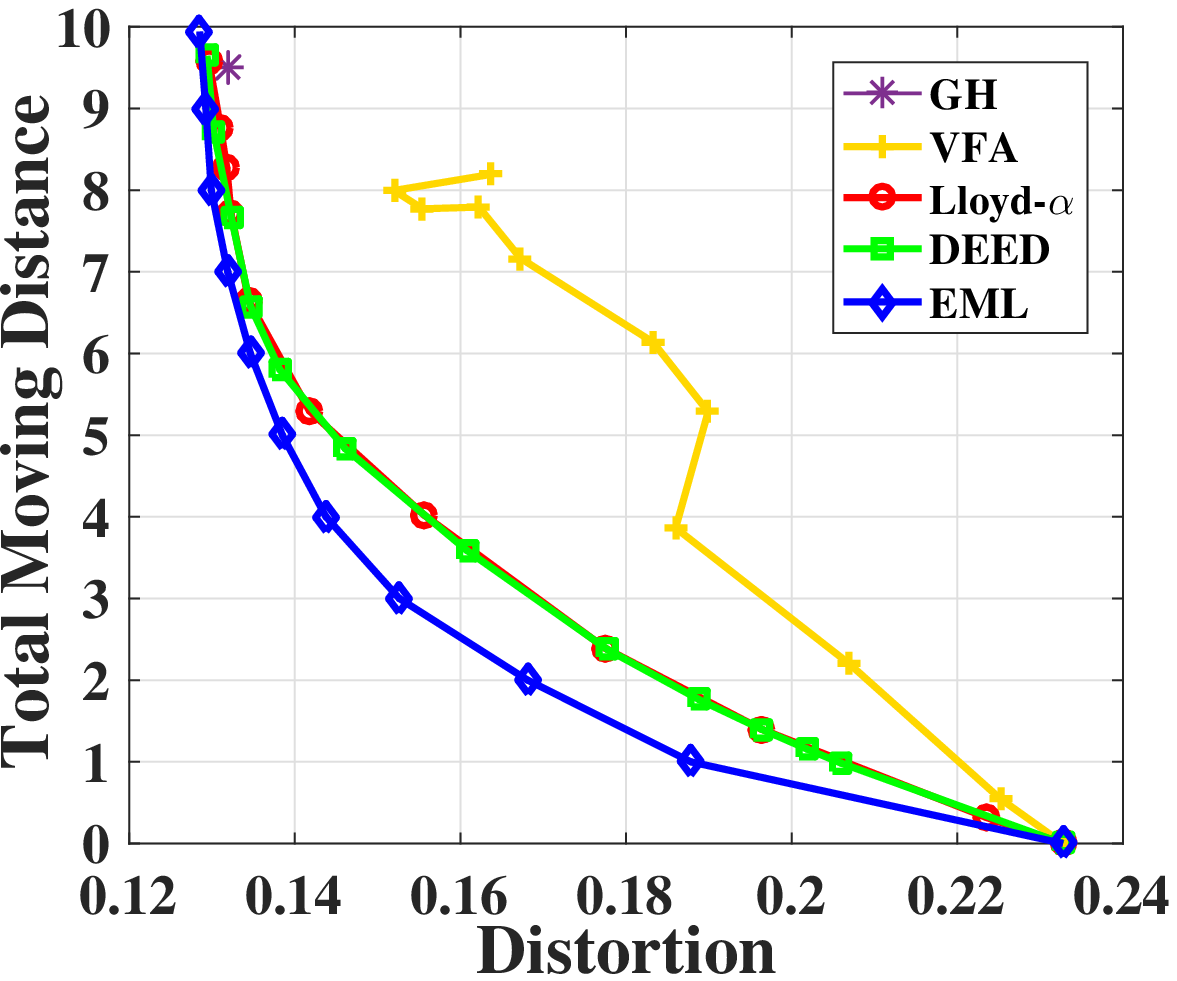}
\label{AreaComparea}}
\hfil
\subfloat[]{\includegraphics[width=3in]{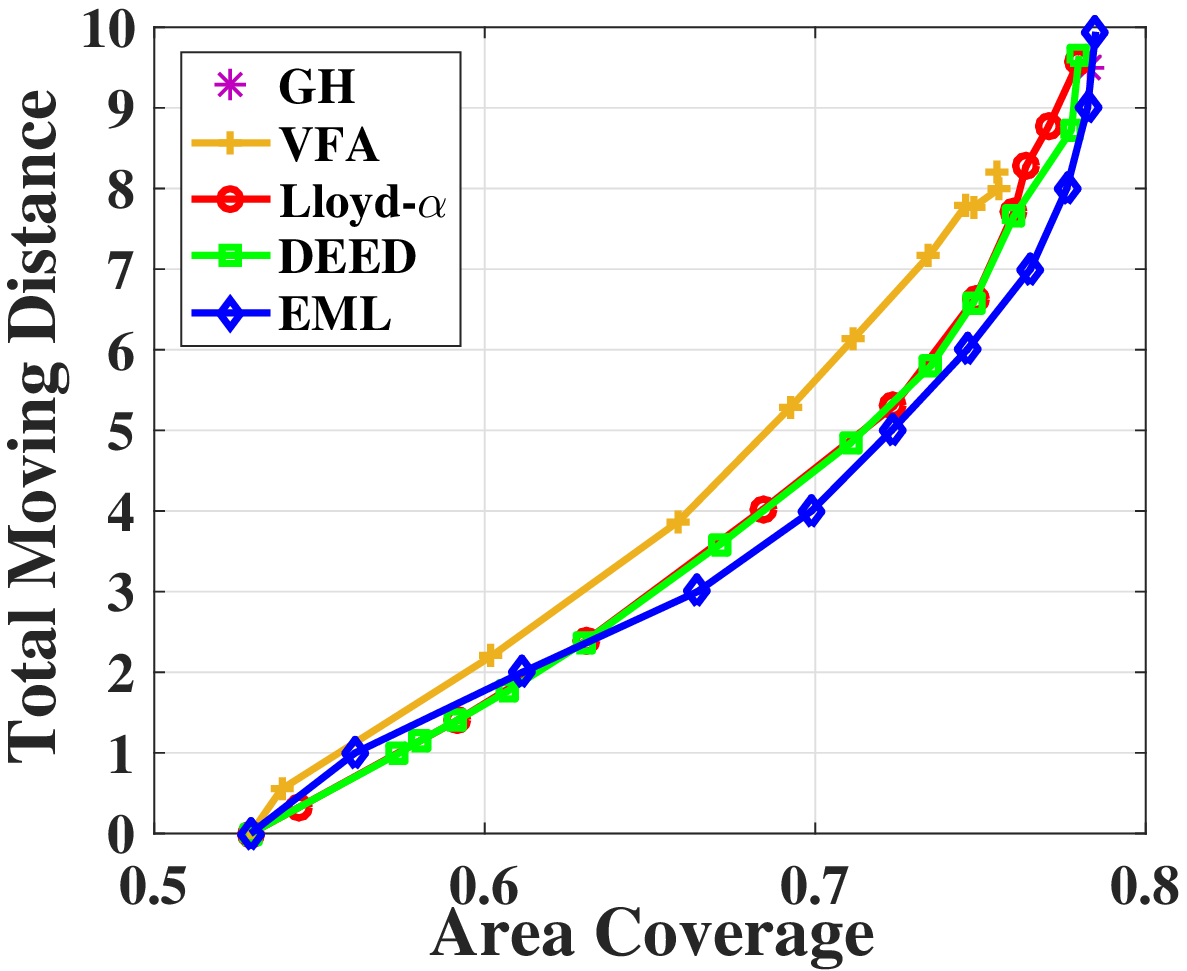}
\label{AreaCompareb}}
\caption{The performance comparison for the sensor deployment with total moving distance constraints in MWSN1.}
\label{AreaCompare}
\end{figure}

\begin{figure}[!t]
\centering
\subfloat[]{\includegraphics[width=3in]{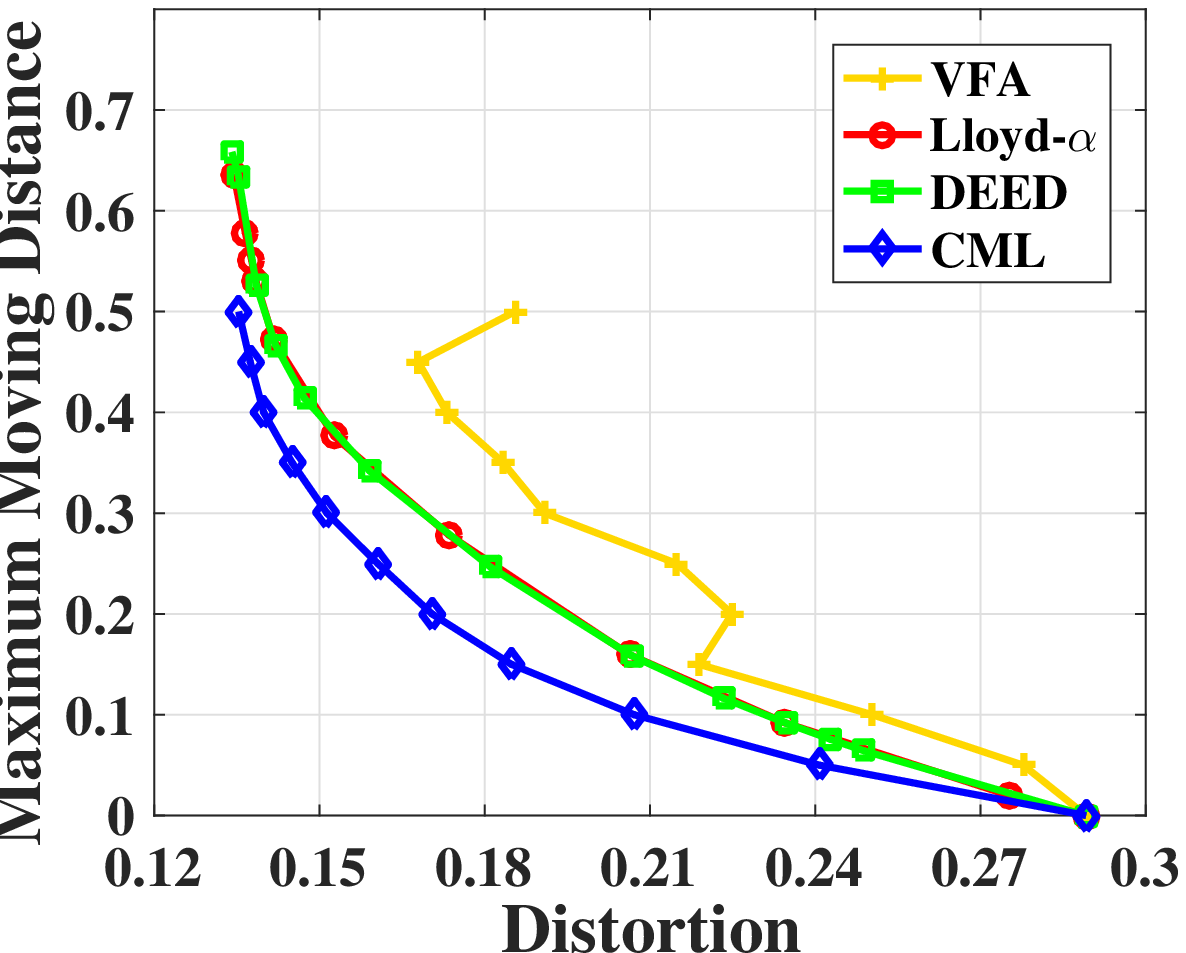}
\label{AreaComparec}}
\hfil
\subfloat[]{\includegraphics[width=3in]{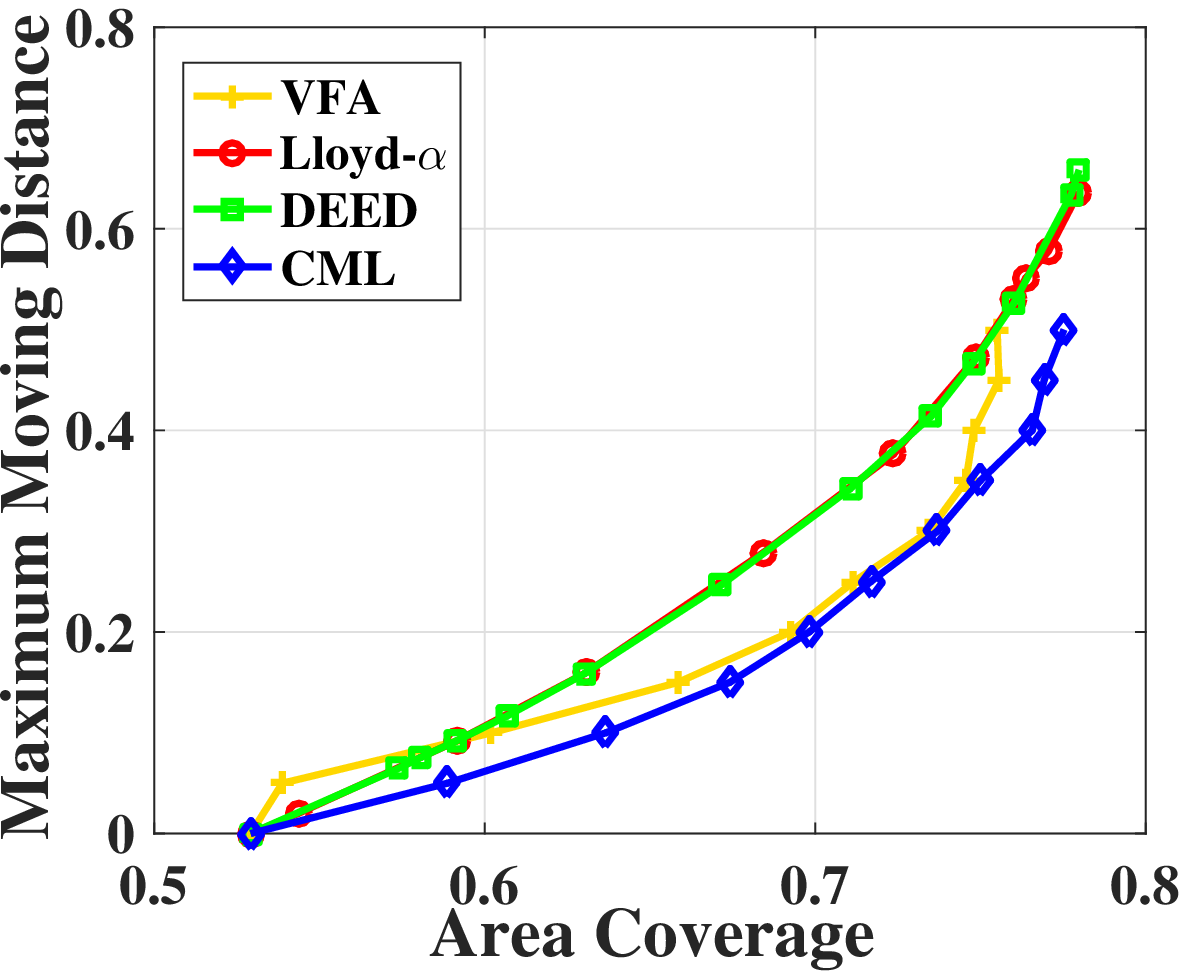}
\label{AreaCompared}}
\caption{The performance comparison for the sensor deployment with maximum individual moving distance constraints in MWSN1.}
\label{AreaCompare2}
\end{figure}

Figs. \ref{AreaCompare} and \ref{AreaCompare2} compare the performance of different algorithms in MWSN1.
Given a total moving distance, EML Algorithm obtains a smaller distortion (and a larger coverage) compared with existing algorithms in the literature.
On the other hand, given a required distortion or coverage, EML Algorithm needs less moving distance, indicating less energy consumption.
Similarly, CML Algorithm can achieve larger coverage with less maximum moving distance (or larger network lifetime) compared with existing algorithms in the literature.
The same conclusion holds in MWSN2. The detailed simulation results in MWSN2 are omitted here to save space.

Furthermore, compared with Lloyd-$\alpha$ and DEED algorithms in \cite{YBZ}, EML and CML algorithms have following advantages.
First, the performance of Lloyd-$\alpha$ and DEED is based on several adjustable parameters, i.e., $\alpha$ and $\delta$.
Unfortunately, there is no explicit relationship between energy consumption and these parameters.
Therefore, to achieve the sensing task with the specific total energy consumption or network lifetime, Lloyd-$\alpha$ and DEED must search different parameters.
On the contrary, the parameters of EML and CML are directly related to the total and individual energy consumption.
One can apply EML or CML to guarantee the specific total energy consumption or network lifetime by simply choosing $\gamma$ or $\gamma_n$.
In addition, our algorithms can be applied to heterogenous MWSNs.
However, the lack of second order derivative of heterogeneous sensing uncertainty prohibits DEED from extending to heterogeneous MWSNs.
\setlength{\floatsep}{10pt}
\begin{figure}[!t]
\centering
\subfloat[]{\includegraphics[width=3in]{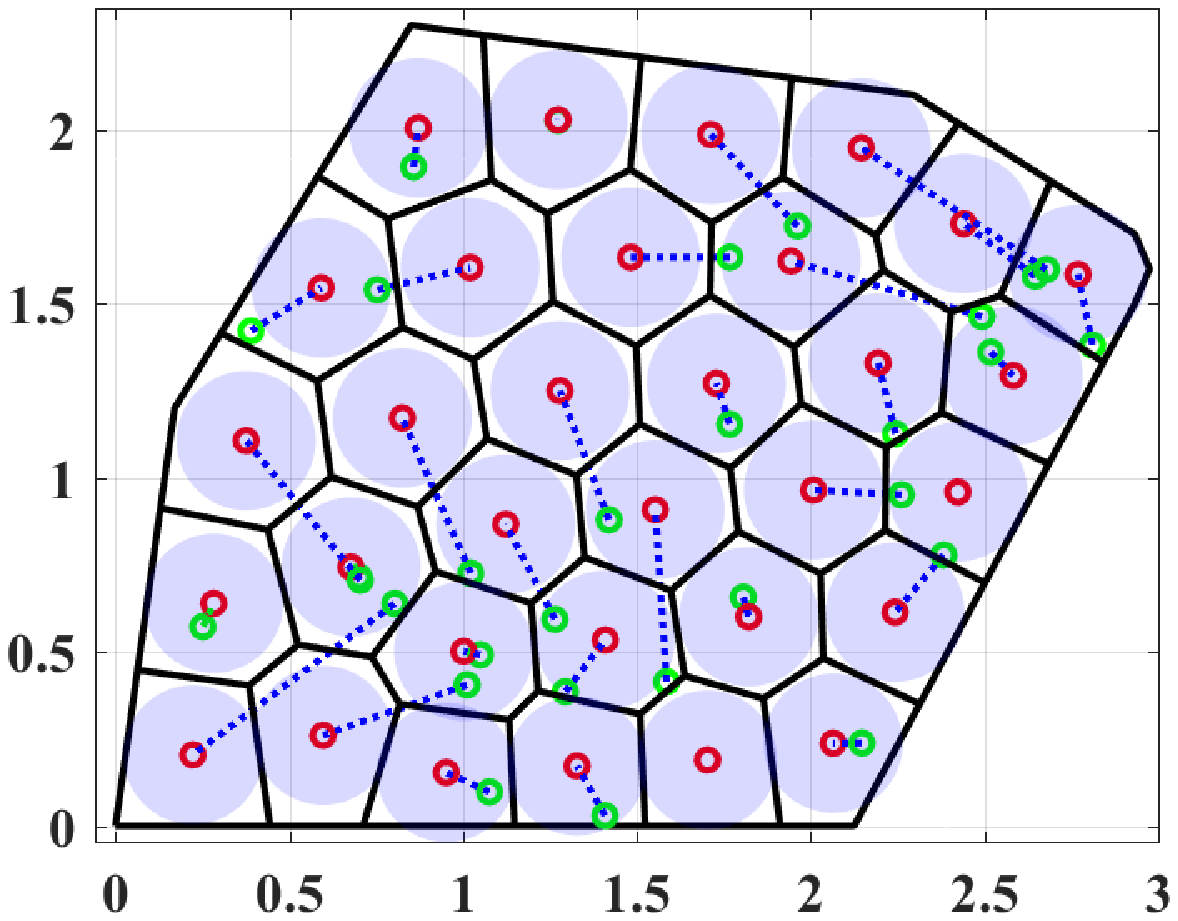}
\label{EML1a}}
\hfil
\subfloat[]{\includegraphics[width=3in]{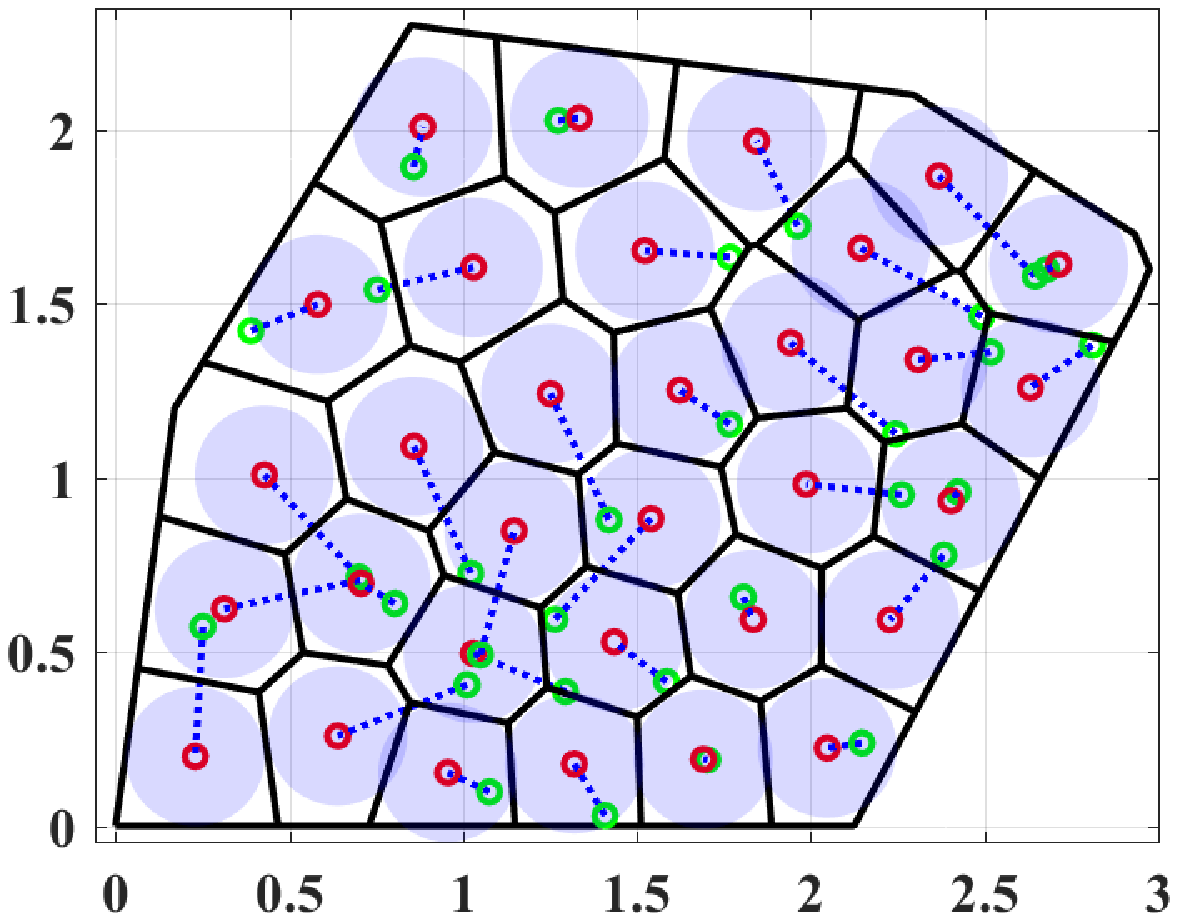}
\label{EML1b}}
\hfil
\subfloat[]{\includegraphics[width=3in]{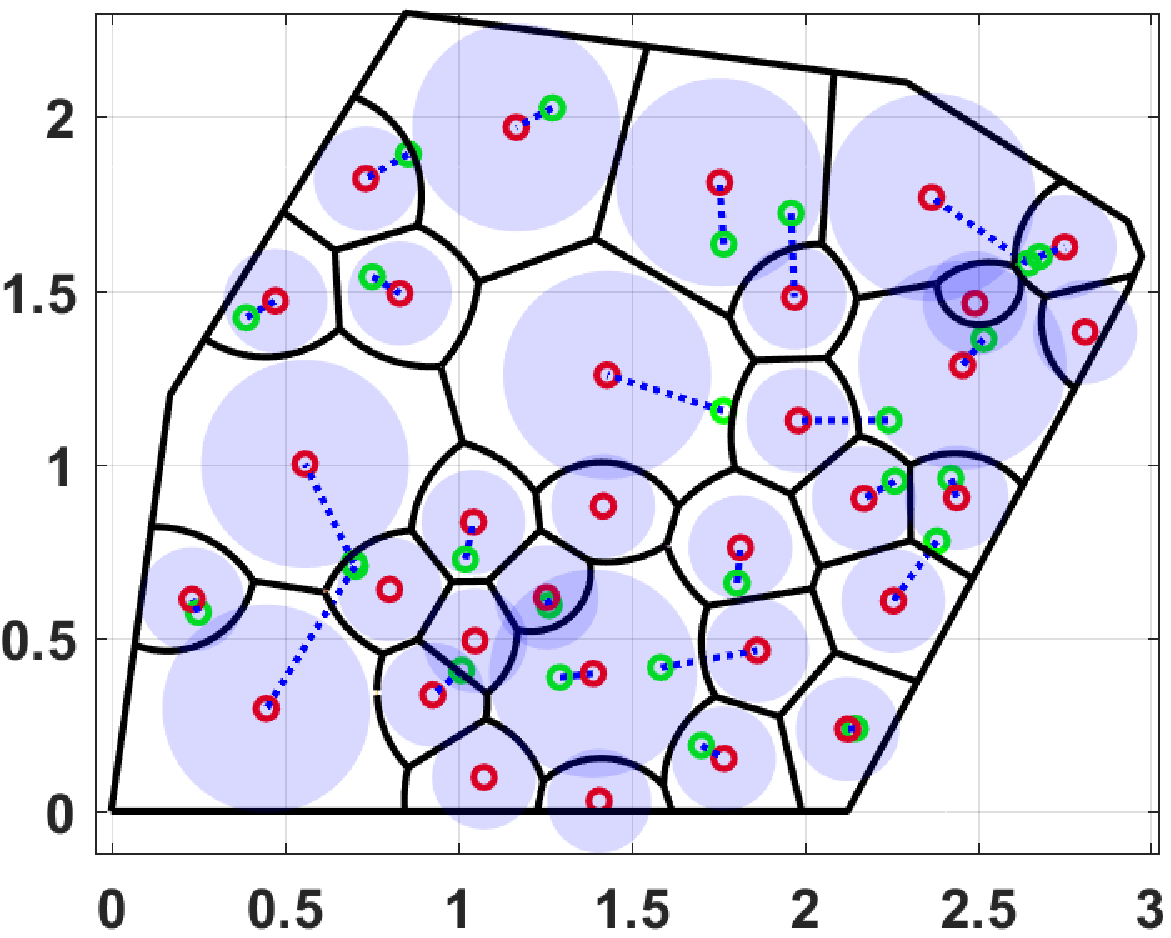}
\label{EML1c}}
\hfil
\subfloat[]{\includegraphics[width=3in]{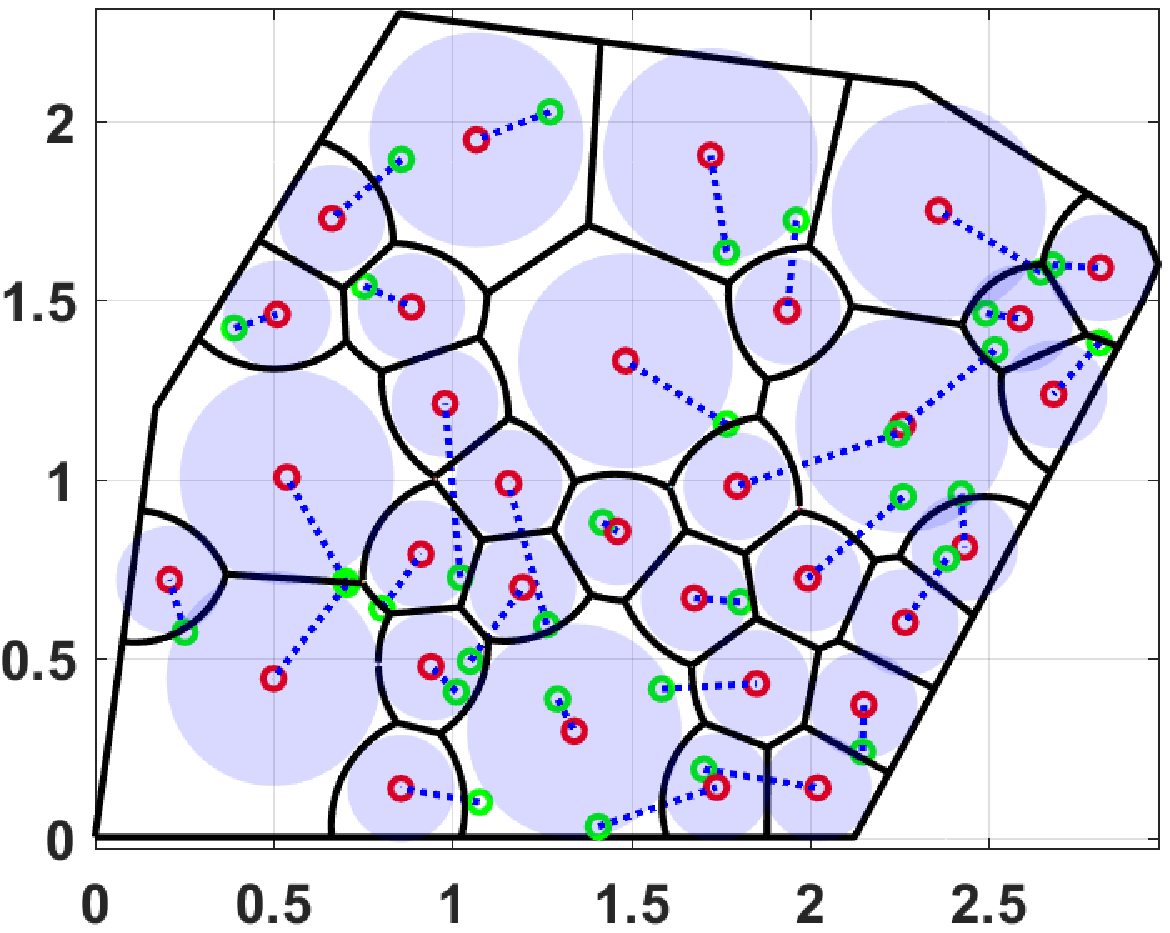}
\label{EML1d}}
\hfil
\captionsetup{justification=justified}
\caption{Sensor deployment: (a) EML in MWSN1; (b) CML in MWSN1; (c) EML in MWSN2; (d) CML in MWSN2. The initial and final sensor locations are, respectively, denoted by green and red circles. The movement paths are denoted by blue lines.}
\label{EML1}
\end{figure}


\section{Conclusions and Discussion}\label{sec:conclusion}
The trade-off between sensing coverage and energy consumption, which is dominated by movement, is discussed in this paper.
We studied the optimal sensor deployment or relocation plan to minimize sensing uncertainty with the total energy constraint.
The necessary condition for optimal deployment implies that some sensors should move towards the centroid and the others should stay put.
Moreover, we discuss the maximum sensing coverage with individual movement constraints, and then propose a necessary condition for the optimal sensor deployment with limited individual energy.
With the help of these necessary conditions, Efficient Movement Lloyd (EML) Algorithm and Constrained Movement Lloyd (CML) Algorithm are designed to obtain the energy-efficient sensor relocation plan.
Our simulation results show that EML and CML algorithms not only achieve significant sensing coverage but also save plenty of energy.

Like area coverage, sensing coverage is defined as the proportion of the target (points or barrier) that covered by at least one sensor.
Intuitively, to decrease the sensing uncertainty, the sensors will be distributed in the area with high density, and thus one can increase the coverage by enlarging the density around the targets.
In other words, the increase in target coverage and barrier coverage, can also be converted to the decrease in the sensing uncertainty with the properly selected density function.
Design of appropriate density functions around point targets and line-shaped barriers is an interesting future work.
Consequently, EML and CML algorithms - which are designed to minimize sensing uncertainty with limited (total or individual) energy - can also be utilized to maximize target and barrier coverage while saving energy.


%
\appendices
\section{Proof of Lemma 1}\label{appendixL1}
Let $\bP^*=\left(p^*_1,\dots,p^*_N\right)$ and $\bR^*=\left(R^*_1,\dots,R^*_N\right)$ be, respectively, the optimal sensor deployment and cell partition.
Since MWVD defined by (\ref{MWVD}) is the optimal cell partition for a given deployment, we have $\bR^*=V(\bP^*)$.
Let $v_n(\bP)=\int_{V_n(\bP)}f(w)dw$ and $c_n(\bP)=\frac{\int_{V_n(\bP)}wf(w)dw}{v_n(\bP)}$ be, respectively, the volume and geometric centroid of the partition $V_n(\bP)$.
The best possible distortion - associated with the optimal sensor deployment $\bP^*$ and cell partition $\bR^*$ - can be rewritten as
\begin{equation}
\begin{aligned}
D(\bP^*) &{=} \sum_{n=1}^{N}\int_{R^*_n}\eta_n\|w-p^*_n\|^2f(w)dw
{=}\sum_{n=1}^{N}\int_{V_n(\bP^*)}\eta_n\|c_n(\bP^*)-w\|^2f(w)dw + \sum_{n=1}^{N} \eta_n\|c_n(\bP^*)-p_n\|^2v_n(\bP^*),
\end{aligned}
\label{L1optD}
\end{equation}
where the second equation follows from the parallel axis theorem.
Now we assume that there exists one sensor $i$ such that its optimal location $p^*_i$ is out of the interval $\overline{\tilde{p}_ic_i(\bP^*)}$.
Let $P'=(p'_1,\dots,p'_N)$ and $P''=(p''_1,\dots,p''_N)$ be two deployments, where $p'_i=c_i(\bP^*)$, $p''_i=\tilde{p}+\|p^*_i-\tilde{p}_i\|\frac{c_i(\bP^*)-\tilde{p}_i}{\|c_i(\bP^*)-\tilde{p}_i\|}$, and $p'_j=p''_j=p^*_j, \forall j\ne i$.

\begin{figure}[!t]
\centering
\subfloat[]{\includegraphics[width=3in]{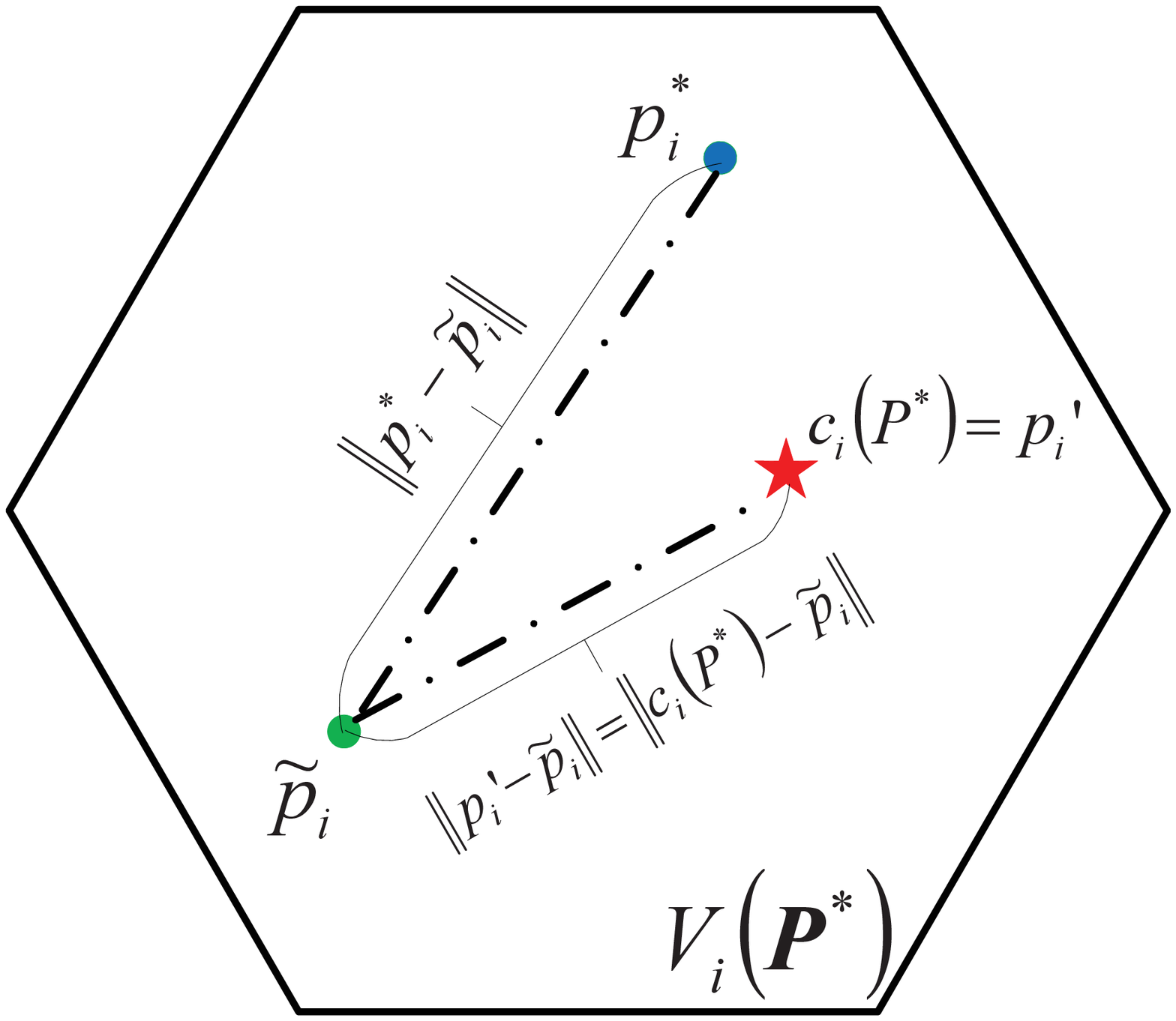}
\label{L1a}}
\hfil
\subfloat[]{\includegraphics[width=3in]{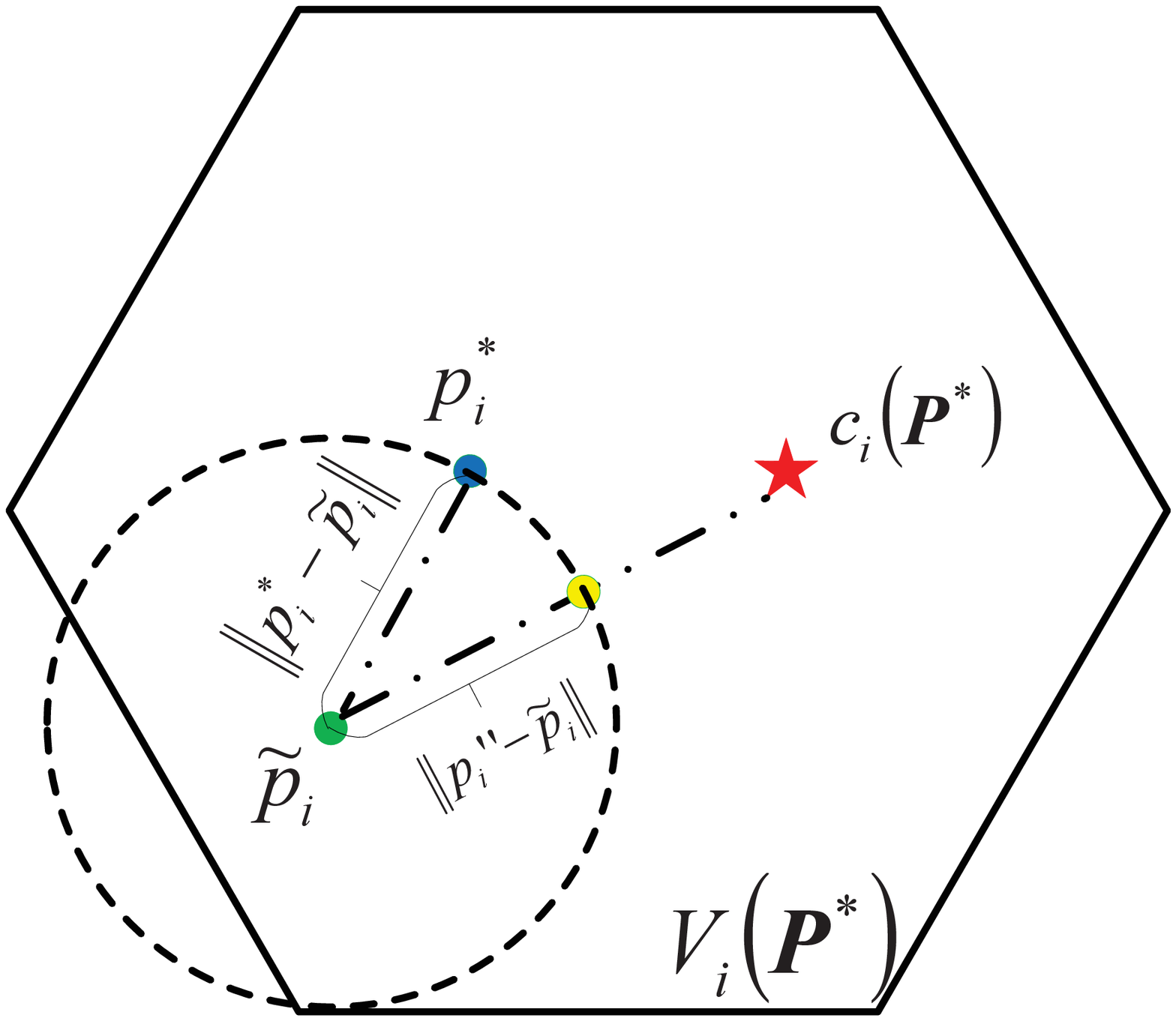}
\label{L1b}}
\caption{An example of sensor relocation in two cases: (a) $\|p^*_i-\tilde{p}_i\|\ge\|c_i(\bP^*)-\tilde{p}_i\|$; (b) $\|p^*_i-\tilde{p}_i\|<\|c_i(\bP^*)-\tilde{p}_i\|$}
\label{L1}
\end{figure}

First, if $\|p^*_i-\tilde{p}_i\|\ge\|c_i(\bP^*)-\tilde{p}_i\|$, it means that we have spent more energy to move to $p^*_i$ compared with moving to the  centroid which is the best possible location.
Obviously, by moving to the centroid, we will have a better distortion and smaller energy consumption and $p^*_i$ cannot be the optimal location.
In what follows, we will prove this fact formally.
The energy consumption for $\bP'$ is
\begin{equation}
\begin{aligned}
E(\bP')
{=}\sum_{n\neq i}\xi_n\|p^*_n-\tilde{p}_n\| + \left(\xi_i\|c_i(\bP^*)-\tilde{p}_i\|\right)
{\le}\sum_{n\neq i}\xi_n\|p^*_n-\tilde{p}_n\| + \left(\xi_i\|p^*_i-\tilde{p}_i\|\right)
{=}E(\bP^*)\le\gamma,
\end{aligned}
\end{equation}
indicating that $\bP'$ is a feasible deployment, see Fig. \ref{L1a}.
Moreover, Since $p^*_i$ is out of the interval $\overline{\tilde{p}_ic_i(\bP^*)}$, the distance between $p^*_i$ and $c_i(\bP^*)$ is positive, i.e., $\|p^*_i-c_i(\bP^*)\|>0$.
Thus, we have
\begin{equation}
\begin{aligned}
&D(\bP^*)-D(\bP'){=}\left[\sum_{n=1}^{N} \eta_n\|c_n(\bP^*)-p^*_n\|^2v_n(\bP^*)\right]-\left[\sum_{n=1}^{N} \eta_n\|c_n(\bP^*)-p'_n\|^2v_n(\bP^*)\right]\\
{=}&\left[\sum_{n=1}^{N} \eta_n\|c_n(\bP^*)-p^*_n\|^2v_n(\bP^*)\right]-\left[\sum_{n\ne i} \eta_n\|c_n(\bP^*)-p^*_n\|^2v_n(\bP^*) + \eta_i\|c_i(\bP^*)-c_i(\bP^*)\|^2v_n(\bP^*)\right]\\
{=}& \eta_i\|c_i(\bP^*)-p^*_i\|^2v_i(\bP^*)>0,
\end{aligned}
\end{equation}
where the first equality follows from (\ref{L1optD}), the second equality follows from $\bP'=(p^*_1,\dots,c_i(\bP^*),\dots,p^*_N)$, and the inequality follows from $\|p^*_i-c_i(\bP^*)\|>0$ and $v_i(\bP^*)>0$.
Therefore, if $\|p^*_i-\tilde{p}_i\|\ge\|c_i(\bP^*)-\tilde{p}_i\|$, $\bP'$ provides a smaller distortion than that of $\bP^*$ without breaking the total energy constraint.

Next, we discuss the case of $\|p^*_i-\tilde{p}_i\|<\|c_i(\bP^*)-\tilde{p}_i\|$, see Fig. \ref{L1b}.
Since $\|p^*_i-\tilde{p}_i\|\in[0,\|c_i(\bP^*)-\tilde{p}_i\|)$, $p''_i=\tilde{p}+\|p^*_i-\tilde{p}_i\|\frac{c_i(\bP^*)-\tilde{p}_i}{\|c_i(\bP^*)-\tilde{p}_i\|}$ is a point on the interval $\overline{\tilde{p}_ic_i(\bP^*)}$ with moving distance
\begin{equation}
\|p''_i-\tilde{p}_i\|=\|p^*_i-\tilde{p}_i\|.
\label{samedist}
\end{equation}
After straightforward calculation, we have $E(\bP'')=\sum_{n=1}^{N}\xi_n\|p''_n-\tilde{p}_n\|=\sum_{n=1}^{N}\xi_n\|p^*_n-\tilde{p}_n\|\le\gamma$, indicating that $\bP''$ is a feasible deployment.
In addition, the difference between the distortions generating from $\bP^*$ and $\bP''$ is
\begin{equation}
\begin{aligned}
&D(\bP^*)-D(\bP''){=}\left[\sum_{n=1}^{N} \eta_n\|c_n(\bP^*)-p^*_n\|^2v_n(\bP^*)\right]-\left[\sum_{n=1}^{N} \eta_n\|c_n(\bP^*)-p''_n\|^2v_n(\bP^*)\right]\\
{=}&\eta_i\|c_i(\bP^*\!)-p^*_i\|^2v_i(\bP^*)-\eta_i\|c_i(\bP^*)-p''_i\|^2v_i(\bP^*)\\
{=}&\eta_iv_i(\bP^*\!)\left(\|c_i(\bP^*\!)\!-\!p^*_i\|\!+\!\|c_i(\bP^*\!)\!-\!p''_i\|\right)\left(\|c_i(\bP^*\!)\!-\!p^*_i\|\!-\!\|c_i(\bP^*\!)-p''_i\|\right)\\
{=}&\eta_iv_i(\bP^*\!)\left(\|c_i(\bP^*\!)\!-\!p^*_i\|\!+\!\|c_i(\bP^*\!)\!-\!p''_i\|\right)\left(\|c_i(\bP^*\!)\!-\!p^*_i\|\!-\!\|c_i(\bP^*\!)-p''_i\|\!+\!\|\tilde{p}_i\!-\!p^*_i\|\!-\!\|\tilde{p}_i-p''_i\|\right)\\
{=}&\eta_iv_i(\bP^*\!)\left(\|c_i(\bP^*\!)\!-\!p^*_i\|\!+\!\|c_i(\bP^*\!)\!-\!p''_i\|\right)\left(\|c_i(\bP^*\!)\!-\!p^*_i\|\!+\!\|\tilde{p}_i\!-\!p^*_i\|\!-\!\|c_i(\bP^*)\!-\!\tilde{p}_i\|\right){>}0,
\end{aligned}
\end{equation}
where the third equality follows from (\ref{samedist}), the fourth equality follows from the condition that $p''_i$ is placed between $\tilde{p}_i$ and $c_i(\bP^*)$, and the inequality follows from the condition that the sum of any two sides of a triangle is greater than the third side.
Therefore, if $\|p^*_i-\tilde{p}_i\|<\|c_i(\bP^*)-\tilde{p}_i\|$, $\bP''$ provides a smaller distortion than that of $\bP^*$ without breaking the total energy constraint.
In sum, one can find a better deployment (either $\bP'$ or $\bP''$), which contradicts our assumption.
\vspace{10pt}
\section{Proof of Proposition 1}\label{appendixP1}
Let $\bP^*=\left(p^*_1,\dots,p^*_N\right)$ and $\bR^*=\left(R^*_1,\dots,R^*_N\right)$ be, respectively, the optimal sensor deployment and cell partition.
Let $v^*_n=\int_{\R_n^*}f(w)dw$ and $c^*_n=\frac{\int_{R_n^*}wf(w)dw}{v^*_n}$ be, respectively, the volume and geometric centroid of the partition $\bR^*$.
The distortion-that is associated with the optimal sensor deployment and cell partition-can be rewritten as
\begin{equation}
\begin{aligned}
D(\bP^*) &{=} \sum_{n=1}^{N}\int_{\bR^*}\eta_n\|w-p^*_n\|^2f(w)dw=\sum_{n=1}^{N}\int_{\bR^*}\eta_n\|c^*_n-w\|^2f(w)dw + \sum_{n=1}^{N} \eta_n\|c^*_n-p^*_n\|^2v^*_n,
\end{aligned}
\label{P1optD}
\end{equation}
where the second equation follows from the parallel axis theorem.
When partition is fixed as $\bR^*$, the first term in (\ref{P1optD}) is a constant.
Thus, $\bP^*$ should be a minimizer of the second term with the constraint (\ref{totalConstraint}), i.e.,
\begin{equation}
\bP^*=\arg\min_{\bP:E(\bP)\le\gamma}\sum_{n=1}^{N} \eta_n\|c^*_n-p_n\|^2v^*_n.
\end{equation}
Obviously, $\widehat{\bP}=(c^*_1,\dots,c^*_N)$ is the minimizer of $F(\bP)=\sum_{n\in\mathcal{I}_{\Omega}} \eta_n\|c^*_n-p_n\|^2v^*_n$ without constraints.
When $E(\widehat{\bP})=\sum_{n\in\mathcal{I}_{\Omega}}\xi_n\|\tilde{p}_n-c^*_n\|=\sum_{n=1}^{N}\Gamma_n(\bP)\le\gamma$, the deployment $\widehat{\bP}$ follows the total energy constraint, and thus $\bP^*=\widehat{\bP}$.
Since MWVD defined by (\ref{MWVD}) is the optimal partition for a given deployment, we have $\bR^*=V(\bP^*)$ and $c^*_n=c_n(\bP^*)$. Replacing $c^*$ by $c_n(\bP^*)$, we get $p^*_n=c_n(\bP^*)$, $\forall n\in\mathcal{I}_{\Omega}$. The necessary condition for the optimal deployment in the case of $\sum_{n=1}^{N}\Gamma_n(\bP)\le\gamma$ is proved.

In what follows, we verify the necessary conditions when $E(\widehat{\bP})=\sum_{n=1}^{N}\Gamma_n(\bP)>\gamma$.
Note that there is not enough energy to move sensors to $\widehat{\bP}$ when $E(\widehat{\bP})>\gamma$.
Under this circumstance, sensors will run out of energy to minimize the distortion, i.e.,
\begin{equation}
\sum_{n\in\mathcal{I}_{\Omega}}\xi_n\|\tilde{p}_n-p^*_n\|=\gamma.
\label{usedout}
\end{equation}
Otherwise, one could further decrease the distortion by moving sensors towards $\widehat{\bP}$.
Now, we start to prove Condition (i).
Suppose there exists one arbitrary sensor $i\in\mathcal{I}_{\Omega}$ and one dynamic sensor $j\in\mathcal{I}_{d}(\bP^*)$ such that $\rho_i(\bP^*)>\rho_j(\bP^*)$.
Let $\Delta=\rho_i(\bP^*)-\rho_j(\bP^*)$ be the difference between the two sensors' moving efficiencies, and $\epsilon\in\left(0,\min\left(\frac{\|p^*_i-c^*_i\|}{\xi_j},\frac{\|p^*_j-\tilde{p}_j\|}{\xi_i},\frac{2\xi_i\xi_j\Delta}{\eta_iv^*_i\xi_j^2+\eta_jv^*_j\xi_i^2}\right)\right)$ be a small value.
By Lemma \ref{oninterval}, $p^*_i$ and $p^*_j$ are, respectively, placed on the segments $\overline{\tilde{p}_ic^*_i}$ and $\overline{\tilde{p}_jc^*_j}$.
Then we have
\begin{equation}
\|\tilde{p}_i-c^*_i\|=\|\tilde{p}_i-p^*_i\|+\|p^*_i-c^*_i\|
\label{pi1}
\end{equation}
and
\begin{equation}
\|\tilde{p}_j-c^*_j\|=\|\tilde{p}_j-p^*_j\|+\|p^*_j-c^*_j\|.
\label{pj1}
\end{equation}
Let $p'_i=\tilde{p}_i+d'_i\cdot\frac{\left(c^*_i-\tilde{p}_i\right)}{\|c^*_i-\tilde{p}_i\|}$ and $p'_j=\tilde{p}_j+d'_j\cdot\frac{\left(c^*_j-\tilde{p}_j\right)}{\|c^*_j-\tilde{p}_j\|}$,
where $d'_i=\left(\|\tilde{p}_i-p^*_i\|+\xi_j\epsilon\right)$ and $d'_j=\left(\|\tilde{p}_j-p^*_j\|-\xi_i\epsilon\right)$.
According to (\ref{pi1}), (\ref{pj1}), and the range of $\epsilon$, we have $d'_i\in\left(\|\tilde{p}_i-p^*_i\|,\|\tilde{p}_i-c^*_i\|\right)$ and $d'_j\in\left(0,\|\tilde{p}_j-p^*_j\|\right)$, indicating that $p'_i$ and $p'_j$ are placed on the intervals $\overline{\tilde{p}_ic^*_i}$ and $\overline{\tilde{p}_jc^*_j}$, respectively.
Thus, we have
\begin{equation}
\|\tilde{p}_i-c^*_i\|=\|\tilde{p}_i-p'_i\|+\|p'_i-c^*_i\|=d'_i+\|p'_i-c^*_i\|=\left(\|\tilde{p}_i-p^*_i\|+\xi_j\epsilon\right)+\|p'_i-c^*_i\|
\label{pi2}
\end{equation}
and
\begin{equation}
\|\tilde{p}_j-c^*_j\|=\|\tilde{p}_j-p'_j\|+\|p'_j-c^*_j\|=d'_j+\|p'_j-c^*_j\|=\left(\|\tilde{p}_j-p^*_j\|-\xi_i\epsilon\right)+\|p'_j-c^*_j\|.
\label{pj2}
\end{equation}
Substituting (\ref{pi1}) to (\ref{pi2}), we get
\begin{equation}
\|p'_i-c^*_i\|=\|p^*_i-c^*_i\|-\xi_j\epsilon.
\label{pi3}
\end{equation}
Substituting (\ref{pj1}) to (\ref{pj2}), we get
\begin{equation}
\|p'_j-c^*_j\|=\|p^*_j-c^*_j\|+\xi_i\epsilon.
\label{pj3}
\end{equation}
Replacing $p^*_i$ and $p^*_j$ in $\bP^*$ by $p'_i$ and $p'_j$, we get another deployment $\bP'=\{p^*_1,\dots,p'_i,\dots,p'_j,\dots,p^*_N\}$.
By straightforward calculation, we have $E(\bP^*)=E(\bP')=\gamma$, indicating that $\bP'$ is a feasible deployment.
Moreover, the difference between $D(\bP^*)$ and $D(\bP')$ is
\begin{align}
&D(\bP^*)-D(\bP')=\sum_{n=1}^{N}\int_{V_n(\bP^*)}\eta_n\|p^*_n-w\|f(w)dw-\sum_{n=1}^{N}\int_{V_n(\bP')}\eta_n\|p'_n-w\|f(w)dw\\
{=}&\int_{\!V\!_n(\bP^*\!)}\!\!\!\!\!\!\!\!\eta_i\|p^*_i\!-\!w\|f(w)dw\!+\!\int_{\!V\!_n(\bP^*\!)}\!\!\!\!\!\!\!\!\eta_j\|p^*_j\!-\!w\|f(w)dw\!-\!\int_{\!V\!_n(\bP'\!)}\!\!\!\!\!\!\!\!\eta_i\|p'_i\!-\!w\|f(w)dw\!-\!\int_{\!V\!_n(\bP'\!)}\!\!\!\!\!\!\!\!\eta_j\|p'_j\!-\!w\|f(w)dw\label{eq1}\\
{=}&\eta_iv^*_i\|p^*_i-c^*_i\|^2+\eta_jv^*_j\|p^*_j-c^*_j\|^2-\eta_iv^*_i\|p'_i-c^*_i\|^2-\eta_jv^*_j\|p'_j-c^*_j\|^2\label{eq2}\\
{=}&\eta_iv^*_i\|p^*_i-c^*_i\|^{\!2}\!\!+\!\eta_jv^*_j\|p^*_j-c^*_j\|^{\!2}\!\!-\!\eta_iv^*_i\left(\|p^*_i-c^*_i\|-\xi_j\epsilon\right)^{\!2}\!\!-\!\eta_jv^*_j\left(\|p^*_j-c^*_j\|+\xi_i\epsilon\right)^{\!2}\label{eq3}\\
{=}&\left[2\eta_iv^*_i\xi_j\|p^*_i-c^*_i\|-2\eta_jv^*_j\xi_i\|p^*_j-c^*_j\|-\left(\eta_iv^*_i\xi_j^2+\eta_jv^*_j\xi_i^2\right)\epsilon\right]\epsilon\label{eq4}\\
{>}&0\label{eq5},
\end{align}
where (\ref{eq1}) follows from parallel axis theorem, (\ref{eq3}) follows from (\ref{pi3}) and (\ref{pj3}), and (\ref{eq5}) follows from the inequation  $0<\epsilon<\frac{2\xi_i\xi_j\Delta}{\eta_iv^*_i\xi_j^2+\eta_jv^*_j\xi_i^2}$.
Therefore, $\bP'$ provides a smaller distortion than that of $\bP^*$ without exceeding the total energy constraint.
In other words, $\bP'$ is a better deployment compared with the optimal deployment $\bP^*$, indicating a contradiction.
Thus, the previous assumption about moving efficiency is invalid, and we have
\begin{equation}
\rho_i(\bP^*)\le\rho_j(\bP^*), \forall i\in I_{\Omega}=\mathcal{I}_{d}\bigcup\mathcal{I}_{s}, j\in\mathcal{I}_{d}.
\label{rhoij}
\end{equation}
Note that (\ref{rhoij}) holds for any $i$ but only for dynamic $j$s.
If index $i$ is static, we have Conclusion (a):
$\rho_i(\bP^*)\le\rho_j(\bP^*), \forall i\in\mathcal{I}_{s}, j\in\mathcal{I}_{d}$.
If two indices $n_1$ and $n_2$ are both dynamic, by choosing $i=n_1$ and $j=n_2$, we will have $\rho_{n_1}\leq \rho_{n_2}$.
On the other hand, if we pick $i=n_2$ and $j=n_1$, we will have  $\rho_{n_2}\leq \rho_{n1}$.
Therefore, for dynamic nodes $n_1$ and $n_2$, we should have $\rho_{n_1}= \rho_{n_2}$.
We denote this common moving efficiency of dynamic sensors by $\bar{\rho}$.
As a result, we have Conclusion (b): $\rho_{i}=\bar{\rho}, \forall i\in\mathcal{I}_{d}$.
Putting (a) and (b) together, we obtain
\begin{equation}
\rho_j(\bP^*)=\bar{\rho}\ge\rho_i(\bP^*), \forall i\in\mathcal{I}_{s}, j\in\mathcal{I}_{d}.
\end{equation}
Next, we calculate the common moving efficiency of dynamic sensors, $\bar{\rho}$.
Using (\ref{pi1}), dynamic sensors' moving efficiencies can be rewritten as
\begin{equation}
\rho_n(\bP^*)=\frac{\xi_n\|p^*_n-c^*_n\|}{\varsigma_n(\bP^*)}=\frac{\xi_n\left(\|\tilde{p}_n-c^*_n\|-\|p^*_n-\tilde{p}_n\|\right)}{\varsigma_n(\bP^*)}=\bar{\rho},  \forall n\in\mathcal{I}_{d}(\bP^*),
\end{equation}
Thus, the dynamic sensor $n$'s energy consumption is
\begin{equation}
E_n(\bP^*)=\xi_n\|p^*_n-\tilde{p}_n\|=\xi_n\|c^*_n-\tilde{p}_n\|-\bar{\rho}\varsigma_n(\bP^*), \forall n\in\mathcal{I}_{d}(\bP^*),
\label{abc}
\end{equation}
and the total energy consumption is
\begin{equation}
E(\bP^*)=\sum_{n=1}^{N}E_n(\bP^*)=\sum_{n\in\mathcal{I}_{d}(\bP^*)}E_n(\bP^*)=\sum_{n=1}^{N}\left[\xi_n\|c^*_n-\tilde{p}_n\|-\bar{\rho}\varsigma_n(\bP^*)\right].
\label{total}
\end{equation}
Substituting (\ref{total}) to (\ref{usedout}), we get $\bar{\rho}=\bar{\rho}(\bP^*)$, indicating Condition (i).


Last, we verify Condition (ii).
By (\ref{abc}), the dynamic sensors' moving distances are
\begin{equation}
\|p^*_n-\tilde{p}_n\|=\|c^*_n-\tilde{p}_n\|-\frac{\bar{\rho}(\bP^*)\varsigma_n(\bP^*)}{\xi_n}, \forall n\in\mathcal{I}_{d}(\bP^*).
\label{bcd}
\end{equation}
Substituting (\ref{pi1}) to (\ref{bcd}), the distance between $p^*_n$ and $c^*_n$ becomes
\begin{equation}
\|p^*_n-c^*_n\|=\frac{\xi_n\bar{\rho}(\bP^*)}{\eta_nv^*_n}, \forall n\in\mathcal{I}_{d}(\bP^*).
\label{pc}
\end{equation}
By Lemma \ref{oninterval}, dynamic sensors' deployments can be represented as
\begin{equation}
p^*_n=c^*_n+\|p^*_n-c^*_n\|\frac{\tilde{p}_n-c^*_n}{\|\tilde{p}_n-c^*_n\|}, \forall n\in\mathcal{I}_{d}(\bP^*).
\label{cp}
\end{equation}
Substituting (\ref{pc}) to (\ref{cp}), Sensor $n$'s optimal location becomes
\begin{equation}
p^*_n=c^*_n+\frac{\xi_n\bar{\rho}(\bP^*)\left(\tilde{p}_n-c^*_n\right)\!}{\eta_nv^*_n\|\tilde{p}_n-c^*_n\|}, \forall n\!\in\!\mathcal{I}\!_{d}(\bP^*)
\label{last}
\end{equation}
Since the optimal partition is $\bR^*=V(\bP^*)$, the optimal geometric centroid and volume are $c^*_n=c_n(\bP^*)$ and $v^*_n=v_n(\bP^*)$.
Replacing $c^*_n$ and $v^*_n$ by $c_n(\bP^*)$ and $v^*_n=v_n(\bP^*)$ in (\ref{last}), we obtain Condition (ii).
\section{Proof of Theorem 1}\label{appendixT1}
To prove Theorem \ref{T1}, we need the following concepts and Lemmas.
Let $\bP^{k}=\left(p^k_1,\dots,p^k_N\right)$ be the sensor deployment after the $k$-th iteration of EML Algorithm.
In particular, $\bP^0=\tilde{\bP}$ is the initial deployment.
We define the energy allocation in the $k$-th iteration as $\bZ^k=(z^k_1,\dots,z^k_N)$, where $z^k_n=\xi_n\|p^{k}_n-\tilde{p}_n\|$ is Sensor $n$'s energy consumption at the end of the $k$-th iteration, where $k\in\{1,\dots,K\}$ and $K$ is the number of iterations in EML Algorithm.
By multiplicatively weighted Voronoi partition in Step (1), the partition during Step (2) is fixed as $\bV(\bP^{k-1})$.
For convenience, let $c^k_n=\frac{\int_{V_n(\bP^k)}wf(w)dw}{\int_{V_n(\bP^k)}f(w)dw}$ and $v^k_n=\int_{V_n(\bP^k)}f(w)dw$ be the geometric centroid and volume of $\bV(\bP^{k})$, respectively.
Moreover, let $\Gamma^{k}_n=c^{k}_n-\tilde{p}_n$ be the distance between Sensor $n$'s initial location and geometric centroid, $\chi^k_n=\xi_n\|\Gamma^k_n\|$ be the energy consumed by moving Sensor $n$ to the geometric centroid, and $\varsigma^k_n=\varsigma_n(\bP^{k})=\frac{\xi^2_n}{\eta_nv^k_n}$ be a parameter depending on the partition volume.
A series of important auxiliary functions $\hat{\rho}^k_n:\Re^{N}\to\Re, k\in\{1,\dots,K\}, n\in\{1,\dots,N\}$, are defined as $\hat{\rho}^k_n(\bZ)=\frac{\chi^{k}_n-z_n}{\varsigma^{k}_n}$, where $\bZ=\left(z_1,\dots,z_n\right)\in\Re^{N}$.
Different from the function $\rho_n(\bP):\Re^{2N}\to\Re$ defined in Section \ref{sec:ProblemA}, the moving efficiency $\hat{\rho}^k_n$ is determined by the energy allocation $\bZ$ rather than the deployment $\bP$.
\begin{lemma}
Let $\mathcal{I}^k_d$ and $\mathcal{I}^k_s$ be, respectively, the dynamic and static sensor sets at the end of the $k$-th iteration.
For EML Algorithm, we make the following conclusions: (a) $\hat{\rho}^{k-1}_i(\bZ^k)=\hat{\rho}^{k-1}_j(\bZ^k), \forall i,j\in\mathcal{I}^k_{d}, k\in\{1,\dots,K\}$ and (b) $\hat{\rho}^{k-1}_i(\bZ^k)\ge\hat{\rho}^{k-1}_{j}(\bZ^k), \forall i\in\mathcal{I}^k_{d}, j\in\mathcal{I}^k_{s}, k\in\{1,\dots,K\}$.
\label{lemma2}
\end{lemma}
\begin{IEEEproof}
At the end of Step (2), dynamic Sensor $n$'s movement in the $k$-th iteration is $p^k_n-\tilde{p}_n\!=\!\left(\!1\!-\!\frac{\varsigma_n(\bP^{k-1})\bar{\rho}(\bP^{k-1})}{\xi_n\|\Gamma_{\!\!n}(\bP^{k\!-\!1})\|}\!\right)\!\Gamma_{\!\!n}(\bP^{k\!-\!1})$ (see Algorithm \ref{EMLA}), and then its energy consumption is
\begin{equation}
z^k_n = \xi_n\|p^k_n-\tilde{p}_n\| =
\xi_n\cdot\left|\!1\!-\!\frac{\varsigma_n(\bP^{k\!-\!1})\bar{\rho}(\bP^{k\!-\!1})}{\xi_n\|\Gamma_{\!\!n}(\bP^{k\!-\!1})\|}\!\right|\cdot\left\|\Gamma_{\!\!n}(\bP^{k\!-\!1})\right\|
=\left|\chi^{k-1}_n-\frac{\varsigma^{k-1}_n\left[\left(\sum_{i\in\mathcal{I}^{k}_d}\chi^{k-1}_i\right)-\gamma\right]}{\sum_{i\in\mathcal{I}^{k}_d}\varsigma^{k-1}_i}\right|, \forall n\in\mathcal{I}^{k}_d,
\end{equation}
where $\mathcal{I}^k_d$ is the dynamic sensor set determined by the $k$-th iteration (Step (2)) of EML Algorithm.
According to lines $9-13$ in Algorithm \ref{EMLA}, the term among the vertical bars is positive, and thus
\begin{equation}
z^k_n=\chi^{k-1}_n-\frac{\varsigma^{k-1}_n\left[\left(\sum_{i\in\mathcal{I}^{k}_d}\chi^{k-1}_i\right)-\gamma\right]}{\sum_{i\in\mathcal{I}^{k}_d}\varsigma^{k-1}_i}.
\label{absz}
\end{equation}
Substituting (\ref{absz}) to the definition of $\hat{\rho}^{k-1}_n(\bZ^k)$, the dynamic sensors' moving efficiencies at $\bZ^k$ can be rewritten as
\begin{equation}
\begin{aligned}
\hat{\rho}^{k-1}_n(\bZ^k) &{=} \frac{\chi^{k-1}_n-z^k_n}{\varsigma^{k-1}_n}
=\frac{\left[\sum_{i\in\mathcal{I}^k_d}\chi^{k-1}_i\right]-\gamma}{\sum_{i\in\mathcal{I}^k_d}\varsigma^{k-1}_i}, \forall n\in\mathcal{I}^k_d.
\end{aligned}
\label{rhok}
\end{equation}
Note that the moving efficiencies for dynamic sensors at $\bZ^k$ are identical, Conclusion (a) is proved.

In what follows, we work on Conclusion (b).
From (\ref{rhok}), we find that $\hat{\rho}^{k-1}_n(\bZ^k)$ depends on the dynamic sensor set $\mathcal{I}^k_d$.
In Step (2) of EML Algorithm, an (inner) iterative algorithm is designed to determine $\mathcal{I}^k_d$.
Without loss of generality, we assume that $L$ inner iterations are used in the $k$-th iteration to determine the dynamic sensor set.
Let $S^k_l$ be the dynamic sensor set after the $l$-th inner iteration in Step (2), where $k$ is the iteration index of EML Algorithm.
In particular, the initial and final dynamic sensor set in Step (2) are $S^k_0=\mathcal{I}_{\Omega}$ and $S^k_L=\mathcal{I}^k_d$.
According to the description of EML Algorithm, the dynamic sensor set is shrunk from $\mathcal{I}_{\Omega}$ to $\mathcal{I}^k_d$ during Step (2), and we have the relationship
\begin{equation}
\mathcal{I}_{\Omega}=S^k_0\supsetneq S^k_1\supsetneq\cdots\supsetneq S^k_{L-1}\supsetneq S^k_{L}=\mathcal{I}^k_d
\end{equation}
In addition, in the $l$-th inner iteration, $l\in\{1,\dots,L\}$, sensors in the set $\left(S^k_{l-1}-S^k_{l}\right)$ are removed from the dynamic sensor set $S^k_{l-1}$ due to the non-positive energy allocation (see Algorithm \ref{EMLA}).
In other words, for any $j\in\left(S^k_{l-1}-S^k_{l}\right)$, we have
\begin{equation}
z^k_j=\chi^{k-1}_j-\varsigma^{k-1}_j\frac{\left[\left(\sum\limits_{i\in S^k_{l-1}}\chi^{k-1}_i\right)-\gamma\right]}{\left(\sum\limits_{i\in S^{k}_{l-1}}\varsigma^{k-1}_i\right)}\le0.
\label{negativedist}
\end{equation}
Expanding (\ref{negativedist}) results in
\begin{equation}
\varsigma^{k-1}_j\left[\left(\sum_{i\in S^{k}_{l-1}}\chi^{k-1}_i\right)-\gamma\right]\ge\chi^{k-1}_j\left(\sum_{i\in S^{k}_{l-1}}\varsigma^{k-1}_i\right).
\label{negativedist2}
\end{equation}
By calculating the summation over $j\in\left(S^k_{l-1}-S^k_{l}\right)$ on both sides of (\ref{negativedist2}), we get
\begin{equation}
\left(\sum_{j\in \left(S^k_{l-1}-S^k_{l}\right)}\varsigma^{k-1}_j\right)\left[\left(\sum_{i\in S^{k}_{l-1}}\chi^{k-1}_i\right)-\gamma\right]\ge\left(\sum_{j\in\left(S^k_{l-1}-S^k_{l}\right)}\chi^{k-1}_j\right)\left(\sum_{i\in S^k_{l-1}}\varsigma^{k-1}_i\right).
\label{ineq1}
\end{equation}
Let $\tilde{\rho}^k(S)=\frac{\left[\sum_{i\in S}\varrho^{k}_i\right]-\gamma}{\sum_{i\in S}\varsigma^k_i}$ be a map from sensor set $S$ to a real number.
The difference of $\tilde{\rho}^{k-1}(\cdot)$ at two dynamic sensor sets that generated by the adjacent inner iterations $l$ and $(l-1)$ is
\begin{align}
&\tilde{\rho}^{k-1}\left(S^k_{l}\right)-\tilde{\rho}^{k-1}\left(S^k_{l-1}\right)\\
=&\frac{\left[\left(\sum\limits_{i\in S^k_{l}}\chi^{k-1}_i\right)-\gamma\right]}{\left(\sum\limits_{i\in S^k_{l}}\varsigma^{k-1}_i\right)}-\frac{\left[\left(\sum\limits_{i\in S^k_{l-1}}\chi^{k-1}_i\right)-\gamma\right]}{\left(\sum\limits_{i\in S^k_{l}}\varsigma^{k-1}_i\right)}\\
=&\frac{\left(\sum\limits_{i\in S^k_{l-1}}\varsigma^{k-1}_i\right)\left[\left(\sum\limits_{i\in S^k_{l}}\chi^{k-1}_i\right)-\gamma\right]
-\left(\sum\limits_{i\in S^k_{l}}\varsigma^{k-1}_i\right)\left[\left(\sum\limits_{i\in S^k_{l-1}}\chi^{k-1}_i\right)-\gamma\right]}{\left(\sum\limits_{i\in S^k_{l}}\varsigma^{k-1}_i\right)\cdot\left(\sum\limits_{i\in S^k_{l-1}}\varsigma^{k-1}_i\right)}\\
=&\frac{\left(\!\sum\limits_{i\in S^{k}_{l\!-\!1}}\!\!\varsigma^{k\!-\!1}_{\!i}\!\right)\!\!\left[\!\left(\!\sum\limits_{i\in S^{k}_{l\!-\!1}}\!\!\chi^{\!k\!-\!1\!}_{\!i}\!\right)\!-\!\left(\!\sum\limits_{j\in \left(\!S^{k}_{l\!-\!1}\!-\!S^{k}_{l}\!\right)}\!\chi^{\!k\!-\!1}_{\!j}\right)\!-\!\gamma\right]\!
-\!\left[\!\left(\sum\limits_{i\in S^{k}_{l\!-\!1}}\!\varsigma^{k\!-\!1}_i\!\right)\!-\!\left(\!\sum\limits_{j\in \left(S^{k}_{l\!-\!1}\!-\!S^{k}_{l}\!\right)}\!\varsigma^{k-1}_j\!\right)\!\right]\!\left[\!\left(\sum\limits_{i\in S^k_{l-1}}\chi^{k-1}_i\!\right)\!-\!\gamma\right]}{\left(\sum\limits_{i\in S^k_{l}}\varsigma^{k-1}_i\right)\cdot\left(\sum\limits_{i\in S^k_{l-1}}\varsigma^{k-1}_i\right)}\\
=&\frac{\left(\sum\limits_{j\in\left(S^{k}_{l-1}-S^{k}_{l}\right)}\varsigma^{k-1}_j\right)\left(\sum\limits_{i\in S^k_{l-1}}\chi^{k-1}_i-\gamma\right)-\left(\sum\limits_{j\in \left(S^{k}_{l-1}-S^{k}_{l}\right)}\chi^{k-1}_j\right)\left(\sum\limits_{i\in S^k_{l-1}}\varsigma^{k-1}_i\right)}{\left(\sum\limits_{i\in S^k_{l}}\varsigma^{k-1}_i\right)\cdot\left(\sum\limits_{i\in S^k_{l-1}}\varsigma^{k-1}_i\right)}\ge0\label{ineq3}, l\in\{1,\dots,L\},
\end{align}
where the inequality in (\ref{ineq3}) follows from (\ref{ineq1}).
In sum, we have the ordered sequence
\begin{equation}
\tilde{\rho}^{k-1}(\mathcal{I}_{\Omega}) = \tilde{\rho}^{k-1}(S^k_{0})\le\tilde{\rho}^{k-1}(S^k_{1})\le\cdots\le\tilde{\rho}^{k-1}(S^k_{L-1})\le\tilde{\rho}^{k-1}(S^k_{L})= \tilde{\rho}^{k-1}(\mathcal{I}^k_d)=\hat{\rho}^{k-1}_n(\bZ^k), \forall n\in\mathcal{I}^k_d
\label{orderedSeq}
\end{equation}
Let $\tilde{\bZ}^k(l)=\left(\tilde{z}^k_1(l),\dots,\tilde{z}^k_N(l)\right)$ be the tentative energy allocation in the $l$-th inner iteration.
According to Algorithm \ref{EMLA} (Line 9), Sensor $n$'s tentative energy consumption in the $l$-th iteration is
\begin{equation}
\tilde{z}^k_n\left(l\right) = \xi_n\left\|\Gamma_n\left(\bP^{k-1}\right)\right\|-\varsigma_n\left(\bP^{k-1}\right)\frac{\left[\left(\sum\limits_{i\in S^{k}_{l}}\chi_i\left(\bP^{k-1}\right)\right)-\gamma\right]}{\left(\sum\limits_{i\in S^{k}_{l}}\varsigma_i\left(\bP^{k-1}\right)\right)} = \chi^{k-1}_n-\varsigma^{k-1}_n\frac{\left[\left(\sum\limits_{i\in S^{k}_{l}}\chi^{k-1}_i\right)-\gamma\right]}{\left(\sum\limits_{i\in S^{k}_{l}}\varsigma^{k-1}_i\right)}, \forall n\in S^{k}_{l},
\end{equation}
and thus the tentative moving efficiency threshold $\tilde{\rho}^{k-1}(S^k_{l})$ can be rewritten as
\begin{equation}
\tilde{\rho}^{k-1}(S^k_{l}) =\frac{\left[\left(\sum\limits_{i\in S^k_l}\chi^{k-1}_i\right)-\gamma\right]}{\left(\sum\limits_{i\in S^{k}_l}\varsigma^{k-1}_i\right)} = \frac{\left(\chi^{k-1}_n-\tilde{z}^k_n(l)\right)}{\varsigma^{k-1}_n}
, \forall n\in S^k_l.
\end{equation}
Note that sensors $j\in\left(S^k_{l-1}-S^k_{l}\right)$ are removed from the dynamic sensor set in the $l$-th iteration because their tentative energy consumption is negative, i.e., $\tilde{z}^k_j\left(l\right)<0$ (see Lines 10-13 in Algorithm \ref{EMLA}).
As a result, we have
\begin{equation}
\hat{\rho}^{k-1}_j(\bZ^k) = \frac{\chi^{k-1}_n}{\varsigma^{k-1}_n} < \frac{\chi^{k-1}_n-\tilde{z}^k_n(l)}{\varsigma^{k-1}_n} = \tilde{\rho}^{k-1}(S^k_{l})
, \forall j\in\left(S^k_{l-1}-S^k_{l}\right), l\in\{1,\dots,L-1\}.
\label{seq2}
\end{equation}
Combining (\ref{orderedSeq}) with (\ref{seq2}), we get
\begin{equation}
\hat{\rho}^{k-1}_j(\bZ^k) \le \hat{\rho}^{k-1}_i(\bZ^k), \forall i\in\mathcal{I}^k_d, j\in\left(S^k_{l-1}-S^k_{l}\right), l\in\{1,\dots,L-1\}.
\end{equation}
Since the static sensor set consists of the sensors that were removed from the first $L-1$ inner iterations, i.e., $\mathcal{I}^k_s=S^k_0-S^k_L=\bigcup_{l\in\{1,\dots,L-1\}}\left(S^k_{l-1}-S^k_{l}\right)$, we get Conclusion (b).
\end{IEEEproof}
\begin{lemma}
For any iteration $k>0$, the deployment $\bP^k$ in EML Algorithm is the unique minimizer of the distortion with the fixed partition, $\bV(\bP^{k-1})$.
\label{minizer}
\end{lemma}
\begin{IEEEproof}
First, we show that the total energy constrained sensor deployment problem with fixed cell partitions can be converted to a simpler optimization problem.
According to Line $3$ in Algorithm \ref{EMLA}, the cell partition is fixed as $\bV(\bP^{k-1})$ in Step (1).
Therefore, using the parallel axis theorem, the distortion in Step (2) can be rewritten as
\begin{equation}
\sum_{n=1}^N\int_{V_n(\bP^{k-1})}\eta_n\|p_n-w\|^2f(w)dw = \sum_{n=1}^N\int_{V_n(\bP^{k-1})}\eta_n\|c^{k-1}_n-w\|^2f(w)dw+\sum_{n=1}^N\eta_n\|p_{n}-c^{k-1}_n\|^2v^{k-1}_n.
\end{equation}
Note that the first term is a constant, and the second term is determined by the distances $\|p_n-c^{k-1}_n\|, n\in\mathcal{I}_{\Omega}$.
It is easy to prove that, the optimal $p_n$ should be placed between $\tilde{p}_n$ and $c^{k-1}_n$ (the proof is similar to that of Lemma \ref{oninterval} and then omitted).
In what follows, we only consider the sensor locations on the segments $\{\overline{\tilde{p}_nc^{k-1}_n}\}_{n\in\mathcal{I}_{\Omega}}$.
Therefore, Sensor $n$'s location at the end of the $k$-th iteration can be represented as a function of $z^k_n$:
\begin{equation}
p_n(z_n)=\tilde{p}_n+\frac{z_n}{\xi_n}\frac{\Gamma^{k-1}_n}{\|\Gamma^{k-1}_n\|}, \forall n\in\mathcal{I}_{\Omega},
\label{T1Z}
\end{equation}
where $z_n$ is Sensor $n$'s energy consumption.
As a result, the distortion associated with the fixed partition $\bV(\bP^{k-1})$ can be represented as a function of energy allocation $\bZ$, i.e.,
\begin{align}
&\widetilde D(\bZ)=D(\bP(\bZ))
{=}\sum_{n=1}^{N}\int_{V_n(\bP^{k-1})}\eta_n\|c^{k-1}_n-w\|^2f(w)dw+\sum_{n=1}^{N}\eta_n\|p_n(z_n)-c^{k-1}_n\|^2v^{k-1}_n\label{DZ0}\\
{=}&\sum_{n=1}^{N}\int_{V_n(\bP^{k-1})}\eta_n\|c^{k-1}_n-w\|^2f(w)dw+\sum_{n=1}^{N}\eta_n\left\|\tilde{p}_n+\frac{z_n}{\xi_n}\frac{\Gamma^{k-1}_n}{\|\Gamma^{k-1}_n\|}-c^{k-1}_n\right\|^2v^{k-1}_n\label{DZ1}\\
{=}&\sum_{n=1}^{N}\int_{V_n(\bP^{k-1})}\eta_n\|c^{k-1}_n-w\|^2f(w)dw+\sum_{n=1}^{N}\eta_n\left\|\frac{z_n}{\xi_n}\frac{\Gamma^{k-1}_n}{\|\Gamma^{k-1}_n\|}-\Gamma^{k-1}_n\right\|^2v^{k-1}_n\label{DZ2}\\
{=}&H^{k-1} + \sum_{n=1}^{N}\frac{\left(z_n-\chi^{k-1}_n\right)^2}{\varsigma^{k-1}_n}\label{DZ4},
\end{align}
where $H^{k-1}=\sum_{n=1}^{N}\int_{V_n(\bP^{k-1})}\eta_n\|c^{k-1}_n-w\|^2f(w)dw$ is the best possible distortion with $\bV^{k-1}$, (\ref{DZ0}) follows from the parallel axis theorem, (\ref{DZ1}) follows from (\ref{T1Z}), (\ref{DZ2}) follows from $\Gamma^{k-1}_n=c^{k-1}_n-\tilde{p}_n$, and (\ref{DZ4}) follows from straightforward calculation.
Thus, the total energy constrained sensor deployment problem in the $k$-th iteration becomes
\begin{align}
&\underset{\bZ}{\text{minimize}} \;\;\;\;\;\widetilde D(\bZ)\label{DZ}\\
&\text{~~~~s.t.} \;\;\;\;\;\;\;\;\;\left(\sum_{n=1}^{N}z_n\right)\le\gamma,
0\le z_n\le\chi^{k-1}_n, n\in I_{\Omega}\label{DZst}.
\end{align}
Obviously, both the objective function (\ref{DZ}) and the constraints (\ref{DZst}) are convex.
Therefore, there exists a unique minimizer for the distortion with fixed cell partition $\bV^{k-1}$.

When $\gamma\ge\sum_{n=1}^{N}\chi^{k-1}_n$ in Step (2), sensors move to the geometric centroid $c^{k-1}_n$ without breaking the energy constraint, indicating an optimum deployment.
On the contrary, when $\gamma<\sum_{n=1}^{N}\chi^{k-1}_n$ in Step (2), there is not enough energy to move every sensor to its geometric centroid $c^{k-1}_n$.
In this case, to minimize the distortion, sensors will run out of the energy, i.e., $\sum_n\xi_nz_n=\gamma$.
Let $\bP^*=\left(p^*_1,\dots,p^*_N\right)$ and $\bZ^*=\left(z^*_1,\dots,z^*_N\right)$ be the optimal deployment and energy allocation when the partition is fixed as $\bV^{k-1}$.
Suppose that the energy allocation $\bZ^{k}$ in the $k$-th iteration is different from the optimum one $\bZ^*=(z^*_1,\dots,z^*_n)$, i.e., $\bZ^{k}\ne\bZ^*$.
Observing that $\sum_{n=1}^{N}z^k_n=\sum_{n=1}^{N}z^*_n=\gamma>0$, one can find two sensors $i$ and $j$ such that $z^*_i>z^k_i\ge0$ and $z^k_j>z^*_j\ge0$.
Since $z^k_j>0$, Sensor $j$ is a dynamic sensor in the $k$-th iteration, i.e., $j\in\mathcal{I}^k_d$.
By Lemma \ref{lemma2}, we have $\hat{\rho}^{k-1}_i(\bZ^k)\le\hat{\rho}^{k-1}_j(\bZ^k)$, and then
\begin{equation}
\hat{\rho}^{k\!-\!1}_{\!i}(\bZ^*)\!=\!\frac{(\chi^{k-1}_i\!-\!z^*_i)}{\varsigma^{k\!-\!1}_i}\!<\!\frac{(\chi^{k-1}_i\!-\!z^{k}_i)}{\varsigma^{k\!-\!1}_i}\!=\!\hat{\rho}^{k\!-\!1}_i(\bZ^k)\!\le\!
\hat{\rho}^{k\!-\!1}_j(\bZ^k)\!=\!\frac{(\chi^{k-1}_j\!-\!z^k_j)}{\varsigma^{k\!-\!1}_j}<\frac{(\chi^{k-1}_j\!-\!z^*_j)}{\varsigma^{k\!-\!1}_j}\!=\!\hat{\rho}^{k\!-\!1}_j(\bZ^*).
\label{ineq2}
\end{equation}
By (\ref{ineq2}), we know $\rho^{k-1}_j(\bZ^*)-\rho^{k-1}_i(\bZ^*)>0$.
Let $\bZ'=(z'_1,\dots,z'_N)$ be a new energy allocation, where $z'_i=z^*_i-\epsilon$, $z'_j=z^*_j+\epsilon$, $z'_k=z^*_k$ for $k\ne i,j$, and $\epsilon$ is an arbitrary value in $\left(0,\min(z^*_i,z^k_j-z^*_j,\frac{2(\hat{\rho}^{k-1}_j(\bZ^*)-\hat{\rho}^{k-1}_i(\bZ^*))}{\left(\varsigma^{k-1}_i\right)^{-1}+\left(\varsigma^{k-1}_j\right)^{-1}}\right)$.
First, we show that $Z'$ satisfies the total energy constraint (\ref{DZst}).
As the optimal energy allocation in the $k$-th iteration, $\bZ^*$ should satisfy the constraint (\ref{DZst}), i.e., $z^*_n\in[0,\xi_n\|\Gamma^{k-1}_{\!n}\|], \forall n\in\mathcal{I}_{\Omega}$.
In addition, since the sensor movement in the $k$-th iteration is $\|p^k_n-\tilde{p}_n\|\!=\!\left(\!1\!-\!\frac{\varsigma_n(\bP^{k\!-\!1})\bar{\rho}(\bP^{k\!-\!1})}{\xi_n\|\Gamma_{\!\!n}(\bP^{k\!-\!1})\|}\!\right)\Gamma_{\!\!n}(\bP^{k\!-\!1}\!)=\left(\!1\!-\!\frac{\varsigma^{k-1}_n\bar{\rho}(\bP^{k\!-\!1})}{\chi^{k-1}_n}\!\right)\Gamma^{k-1}_{\!\!n}$, we have $z^k_n=\xi_n\|p^k_n-\tilde{p}_n\|\in[0,\chi^{k-1}_n], \forall n\in\mathcal{I}_{\Omega}$.
According to the ranges of $z^*_n$, $z^k_n$, and $\epsilon$, we get $z'_i\in[0,\chi^{k-1}_i]$, $z'_j\in[0,\chi^{k-1}_j]$, and $\sum_{n=1}^{N}z'_n=\gamma$, indicating $\bZ'$ is a feasible energy allocation.
Next, we show that $\bZ'$ provides a smaller distortion compared to $\bZ^*$.
The difference between $\widetilde{D}(\bZ^*)$ and $\widetilde{D}(\hat{\bZ})$ is
\begin{equation}
\begin{aligned}
& \widetilde{D}(\bZ^*)-\widetilde{D}(\hat{\bZ'})
{=} \frac{\left(\chi^{k-1}_i-z^*_i\right)^2}{\varsigma^{k-1}_i}+\frac{\left(\chi^{k-1}_j-z^*_j\right)^2}{\varsigma^{k-1}_j}-\frac{\left(\chi^{k-1}_i-\left(z^*_i-\epsilon\right)\right)^2}{\varsigma^{k-1}_i}-\frac{\left(\chi^{k-1}_j-\left(z^*_j+\epsilon\right)\right)^2}{\varsigma^{k-1}_j}\\
{=}& \epsilon\left[2\left(\hat{\rho}^{k-1}_j(\bZ^*)-\hat{\rho}^{k-1}_i(\bZ^*)\right)-\left(\left(\varsigma^{k-1}_i\right)^{-1}+\left(\varsigma^{k-1}_j\right)^{-1}\right)\epsilon\right]
{>}0,
\end{aligned}
\end{equation}
where the inequality follows from $\epsilon<\frac{2\left(\hat{\rho}^{k-1}_j(\bZ^*)-\hat{\rho}^{k-1}_i(\bZ^*)\right)}{\left(\varsigma^{k-1}_i\right)^{-1}+\left(\varsigma^{k-1}_j\right)^{-1}}$.
In sum, $\bZ'$ is a better solution than $\bZ^*$, which contradicts the initial assumption.
Therefore, $\bZ^k$ is the unique optimal solution for the converted optimized problem defined by (\ref{DZ}) and (\ref{DZst}).
Consequently, $\bP^k$ is the unique minimizer of the distortion with the fixed partition, $\bV(\bP^{k-1})$.
\end{IEEEproof}

Now we have enough materials to prove Theorem \ref{T1}.
EML Algorithm is an iterative improvement algorithm only if both steps in EML do not increase the distortion defined by (\ref{distortion}) subject to the constraint defined by (\ref{totalConstraint}).
In Section \ref{sec:model}, we have proved that MWVD is the optimal cell partition for a given deployment.
Therefore, Step (1) of EML Algorithm will definitely decrease the distortion.
During Step (2) in the $k$-th iteration, the cell partition is fixed as $\bV(\bP^{k-1})$.
By Lemma \ref{minizer}, we know Step (2) in the $k$-th iteration will find the unique minimizer for the distortion with fixed partition $\bV(\bP^{k-1})$.
In other words, the distortion will be decreased by Step (2) in EML Algorithm unless the termination condition, $\bP^{k}=\bP^{k-1}$, happens.
Therefore, EML Algorithm is an iterative improvement algorithm.
In addition, the distortion has a lower bound 0.
As a result, the distortion of EML Algorithm is non-increasing with a lower bound, indicating that the distortion converges.


\section{Proof of Proposition 2}\label{appendixP2}
Let $\bR^*=(R^*_1,\dots,R^*_N)$ be the optimal partition.
For simplicity, let $c^*_n=\frac{\int_{R^*_n}wf(w)dw}{\int_{R^*_n}f(w)dw}$ and $v^*_n=\int_{R^*_n}f(w)dw$ be, respectively, the geometric centroid and the Lebesgue measure (volume) of $R^*_n$, $\forall n\in\mathcal{I}_{\Omega}$.
The minimum distortion can be rewritten as (\ref{P1optD}).
The first term in (\ref{P1optD}) is a constant and thus the second term should be minimized with the constraint (\ref{individualConstraint}).
Obviously, there is no correlation between different sensor locations.
Thus, we can optimize the distortion over each sensor, separately.
Therefore, the optimal sensor location $p^*_n$ should minimize the distance
to $c^*_n$ subject to the constraint that $p_n$ remains within a ball of radius $\frac{\gamma_n}{\xi_n}$ centered at $\tilde{p}_n$.
A simple geometric argument reveals that the optimal solution should fall on the interval $\overline{c^*_n\tilde{p}_n}$.
If the radius $\frac{\gamma_n}{\xi_n}$ is larger than the distance between $c^*_n$ and $\tilde{p}_n$, Sensor $n$ should be placed at $c^*_n$, i.e.,
\begin{equation}
p^*_n=c^*_n, \forall n\in I_{\Omega}.
\label{p1}
\end{equation}
Otherwise, Sensor $n$ should be placed at a point on the interval $\overline{c^*_n\tilde{p}_n}$ which is closest to $c^*_n$, i.e., $\|p^*_n-\tilde{p}\|=\frac{\gamma_n}{\xi_n}$ and then 
\begin{equation}
p^*_n=p_n^{(0)}+\frac{c^*_n-\tilde{p}_n}{\|c^*_n-\tilde{p}_n\|}\frac{\gamma_n}{\xi_n}, \forall n\in I_{\Omega}.
\label{p2}
\end{equation}
Since MWVD is the optimal partition, we have $\bR^*=V(\bP^*)$
and then
\begin{equation}
c^*_n=\frac{\int_{V_n(\bP^*)}wf(w)dw}{\int_{V_n(\bP^*)}f(w)dw}=c_n(\bP^*), \forall n\in I_{\Omega}.
\label{c}
\end{equation}
Substituting (\ref{c}) to (\ref{p1}) and (\ref{p2}), we get the necessary condition (\ref{optP}).
\section{Proof of Theorem 2}\label{appendixT2}
To prove Theorem \ref{T2}, we adopt the concepts defined in Appendix \ref{appendixT1}.
CML Algorithm is an iterative improvement algorithm only if both steps in EML do not increase the distortion defined by (\ref{distortion}) subject to the constraint defined by (\ref{individualConstraint}).
In Section \ref{sec:model}, we have proved that MWVD is the optimal cell partition for a given deployment.
Therefore, Step (1) of CML Algorithm will not increase the distortion.
During the second step of CML's $k$-th iteration, the cell partition is fixed as $\bV(\bP^{k-1})$.
When cell partition is fixed, sensors are independent (there is no interaction among sensors in both objective function (\ref{distortion}) and constraints (\ref{individualConstraint})).
In this case, problem $\mathcal{B}$ can be converted into $N$ independent optimization problems in which the $n$-th optimization problem is
\begin{align}
\underset{p^k_n}{\text{minimize}}  &\int_{V_n(\bP^{k-1})}\|p^k_n-w\|^2f(w)dw \label{locd}\\
\!\!\!\!\!\!\!\text{~~~~s.t.} \;\;\;\;\;\; &\xi_n\|p^k_n-\tilde{p}_n\|\leq\gamma_n\label{localconstraint}.
\end{align}
To follow the constraint (\ref{localconstraint}), Sensor $n$ should move within $B(\tilde{p}_n,\frac{\gamma_n}{\xi_n})$, a disk centered at $\tilde{p}_n$ with radius $\frac{\gamma_n}{\xi_n}$.
In addition, by the parallel axis theorem, sub-objective function (\ref{locd}) can be rewritten as $\int_{V_n(\bP^{k-1})}\|c^{k-1}_n-w\|^2f(w)dw + \|p^k_n-c^{k-1}_n\|^2v^{k-1}_n$, which is determined by the distance from Sensor $n$'s location in the $k$-th iteration, $p^k_n$, to its geometric centroid in the $k-1$th iteration, $c^{k-1}_n$.
It is self-evident that the Sensor $n$'s movement in Step (2) minimizes the distance to $c^{k-1}_n$ within the disk $B(\tilde{p}_n, \frac{\gamma_n}{\xi_n})$.
In other words, the sensor movement in Step (2) minimizes the distortion while following the individual energy constraints (\ref{localconstraint}).
Therefore, CML Algorithm is an iterative improvement algorithm.
In addition, the distortion has a lower bound 0.
As a result, the distortion of CML Algorithm is non-increasing with a lower bound, indicating that the distortion converges.

\begin{table*}[!t]
\centering
\caption{Notation Table
(the symbols with $^*$ means the optimal one)}
\begin{tabular}{|c|c|}
\hline
Symbol & Description\\
\hline
$N$ & The number of sensors\\
\hline
$n$ & Index for sensors\\
\hline
$\Omega$ & Target region\\
\hline
$\tilde{\bP}$ & Initial sensor deployment\\
\hline
$\tilde{p}_n$ & Sensor $n$'s initial deployment\\
\hline
$\bP$ & Sensor deployment\\
\hline
$p_n$ & Sensor $n$' location\\
\hline
$\bP^k$ & Sensor deployment at the end of the $k$-th iteration\\
\hline
$p^k_n$ & Sensor $n$' location at the end of the $k$-th iteration\\
\hline
$\bZ^k$ & Energy allocation at the end of the $k$-th iteration\\
\hline
$z^k_n$ & Sensor $n$' energy consumption at the end of the $k$-th iteration\\
\hline
$\bR$ & Complete cell partition\\
\hline
$R_n$ & Sensor $n$' cell\\
\hline
$f(\cdot)$ & Density function\\
\hline
$\mathcal{I}_{\Omega}$ & Set of sensors\\
\hline
$\mathcal{I}_{d}(\cdot)$ & Set of dynamic sensors\\
\hline
$\mathcal{I}_{s}(\cdot)$ & Set of static sensors\\
\hline
$\mathcal{I}^k_{d}$ & Set of dynamic sensors at the end of the $k$-th iteration\\
\hline
$\mathcal{I}^k_{s}$ & Set of static sensors  at the end of the $k$-th iteration\\
\hline
$D(\cdot)$ & Sensing uncertainty (distortion)\\
\hline
$D_n(\cdot)$ & Sensing $m$'s local distortion\\
\hline
$E(\cdot)$ & Total energy consumption\\
\hline
$E_n(\cdot)$ & Sensor $n$'s energy consumption\\
\hline
$E^{R}$ & Total residual energy\\
\hline
$e_n$ & Sensor $n$'s residual energy\\
\hline
$T$ & Network lifetime\\
\hline
$C^{\mathcal{A}}(\cdot)$ & Area coverage\\
\hline
$\mathbf{card}(\cdot)$ & cardinality\\
\hline
$\|\cdot\|$ & Euclidean distance\\
\hline
$\overline{ab}$ & The interval between points a and b.\\
\hline
$R_s$ & The minimum sensing radius\\
\hline
$\eta_n$ & Sensor $n$'s sensing cost parameter\\
\hline
$\xi_n$ & Sensor $n$'s moving cost parameter\\
\hline
$v_n(\cdot)$ & Lebesgue measure (volume) of the $n$th Voronoi Diagram generated by $\bP$\\
\hline
$c_n(\cdot)$ & Geometric of the $n$-th Voronoi Diagram generated by $\bP$\\
\hline
$\gamma$ & Total energy constraint\\
\hline
$\gamma_n$ & Sensor $n$'s energy constraint\\
\hline
$\rho_n(\bP)$ & Sensor $n$'s moving efficiency at deployment $\bP$\\
\hline
$\bar{\rho}(\bP)$ & Moving efficiency threshold at $\bP$\\
\hline
$\hat{\rho}^k_n(\bZ)$ & Sensor $n$'s moving efficiency with partition $\bV(\bP^{k})$ and energy allocation $\bZ$\\
\hline
$\bar{\rho}^k$ & Moving efficiency threshold at the end of the $k$-th iteration\\
\hline
$\tilde{\rho}^k_n(S)$ & Sensor $n$'s tentative moving efficiency when the dynamic sensor set is $S$\\
\hline
$\Gamma_n(\cdot)$ & An auxiliary parameter that defined by $\Gamma_n(\bP)=\|\tilde{p}_n-c_n(\bP)\|$\\
\hline
$\varrho_n(\cdot)$ & An auxiliary parameter that defined by $\varrho_n(\bP)=\xi_n\|p_n-c_n(\bP)\|$\\
\hline
$\varsigma_n(\cdot)$ & An auxiliary parameter that defined by $\varsigma_n(\bP)=\frac{\xi^2_n}{\eta_nv_n(\bP)}$\\
\hline
$\Gamma^k_n$ & An auxiliary parameter that defined by $\Gamma^k_n=\|\tilde{p}_n-c^{k}_n\|$\\
\hline
$\chi^k_n$ & An auxiliary parameter that defined by $\chi^k_n=\xi_n\|\tilde{p}_n-c^k_n\|$\\
\hline
$\varsigma^k_n$ & An auxiliary parameter that defined by $\varsigma^k_n=\frac{\xi^2_n}{\eta_nv^{k}_n}$\\
\hline
\end{tabular}
\linespread{1}
\label{NotationTable}
\end{table*}

\end{document}